\documentclass[journal]{IEEEtran}
\usepackage{amsmath,amsfonts}
\usepackage{algorithm}
\usepackage{array}
\usepackage{textcomp}
\usepackage{stfloats}
\usepackage{url}
\usepackage{verbatim}
\usepackage{graphicx}
\usepackage{cite}
\usepackage{graphicx}%
\usepackage{multirow}%
\usepackage{amsmath,amssymb,amsfonts}%
\usepackage{amsthm}%
\usepackage{mathrsfs}%
\usepackage{xcolor}%
\usepackage{textcomp}%
\usepackage{manyfoot}%
\usepackage{booktabs}%
\usepackage{algorithm}%
\usepackage{algorithmicx}%
\usepackage{algpseudocode}%
\usepackage{listings}%

\usepackage{nicefrac}       
\usepackage{epsfig}
\usepackage{subfigure}
\usepackage{url}
\usepackage{bm}
\usepackage{footmisc}
\usepackage{booktabs}       
\usepackage{color}
\usepackage{makecell}
\usepackage{hyperref}       
\usepackage{lineno}   
\usepackage{soul}
\usepackage{stfloats} 

\def\l{\left}
\def\r{\right}
\newtheorem*{customlemma}{Theorem}
\newtheorem*{remark}{Remark}
\newtheorem*{notations}{Notations}
\usepackage{bbding}
\usepackage[normalem]{ulem}
\useunder{\uline}{\ul}{}

\begin{document}

\title{RSF-Conv: Rotation-and-Scale Equivariant \\ Fourier Parameterized Convolution for \\ Retinal Vessel Segmentation}
\author{Zihong Sun, Hong Wang, Qi Xie, Yefeng Zheng, \IEEEmembership{Fellow, IEEE}, Deyu Meng, \IEEEmembership{Member, IEEE}
\thanks{\!\!\!\!\!\!\!\!\!\!\!\! Zihong Sun, Qi Xie, and Deyu Meng are with the School of Mathematics and Statistics and Ministry of Education Key Laboratory of Intelligent Networks
and Network Security, Xi’an Jiaotong University, Shaanxi 710049, China.
Email: szhc0gk@stu.xjtu.edu.cn, xie.qi@mail.xjtu.edu.cn, and dymeng@mail.xjtu.edu.cn}
\thanks{\!\!\!\!\!\!\!\!\!\!\!\! Hong Wang and Yefeng Zheng are with the Jarvis Research Center, Tencent YouTu Lab, Shenzhen, China.
Email: hazelhwang@tencent.com and yefengzheng@tencent.com}
}

\markboth{Journal of \LaTeX\ Class Files,~Vol.~14, No.~8, August~2021}%
{Shell \MakeLowercase{\textit{et al.}}: A Sample Article Using IEEEtran.cls for IEEE Journals}


\maketitle

\begin{abstract}
Retinal vessel segmentation is of great clinical significance for the diagnosis of many eye-related diseases, but it is still a formidable challenge due to the intricate vascular morphology. With the skillful characterization of the translation symmetry existing in retinal vessels, convolutional neural networks (CNNs) have achieved great success in retinal vessel segmentation. However, the rotation-and-scale symmetry, as a more widespread image prior in retinal vessels, fails to be characterized by CNNs. Therefore, we propose a rotation-and-scale equivariant Fourier parameterized convolution (RSF-Conv) specifically for retinal vessel segmentation, and provide the corresponding equivariance analysis. As a general module, RSF-Conv can be integrated into existing networks in a plug-and-play manner while significantly reducing the number of parameters. For instance, we replace the traditional convolution filters in U-Net and Iter-Net with RSF-Convs, and faithfully conduct comprehensive experiments. RSF-Conv+U-Net and RSF-Conv+Iter-Net not only have slight advantages under in-domain evaluation, but more importantly, outperform all comparison methods by a significant margin under out-of-domain evaluation. It indicates the remarkable generalization of RSF-Conv, which holds greater practical clinical significance for the prevalent cross-device and cross-hospital challenges in clinical practice. To comprehensively demonstrate the effectiveness of RSF-Conv, we also apply RSF-Conv+U-Net and RSF-Conv+Iter-Net to retinal artery/vein classification and achieve promising performance as well, indicating its clinical application potential.
\end{abstract}


\section{Introduction}
\label{sec:introduction}

\begin{figure}[!t]
\centerline{\includegraphics[width=\columnwidth]{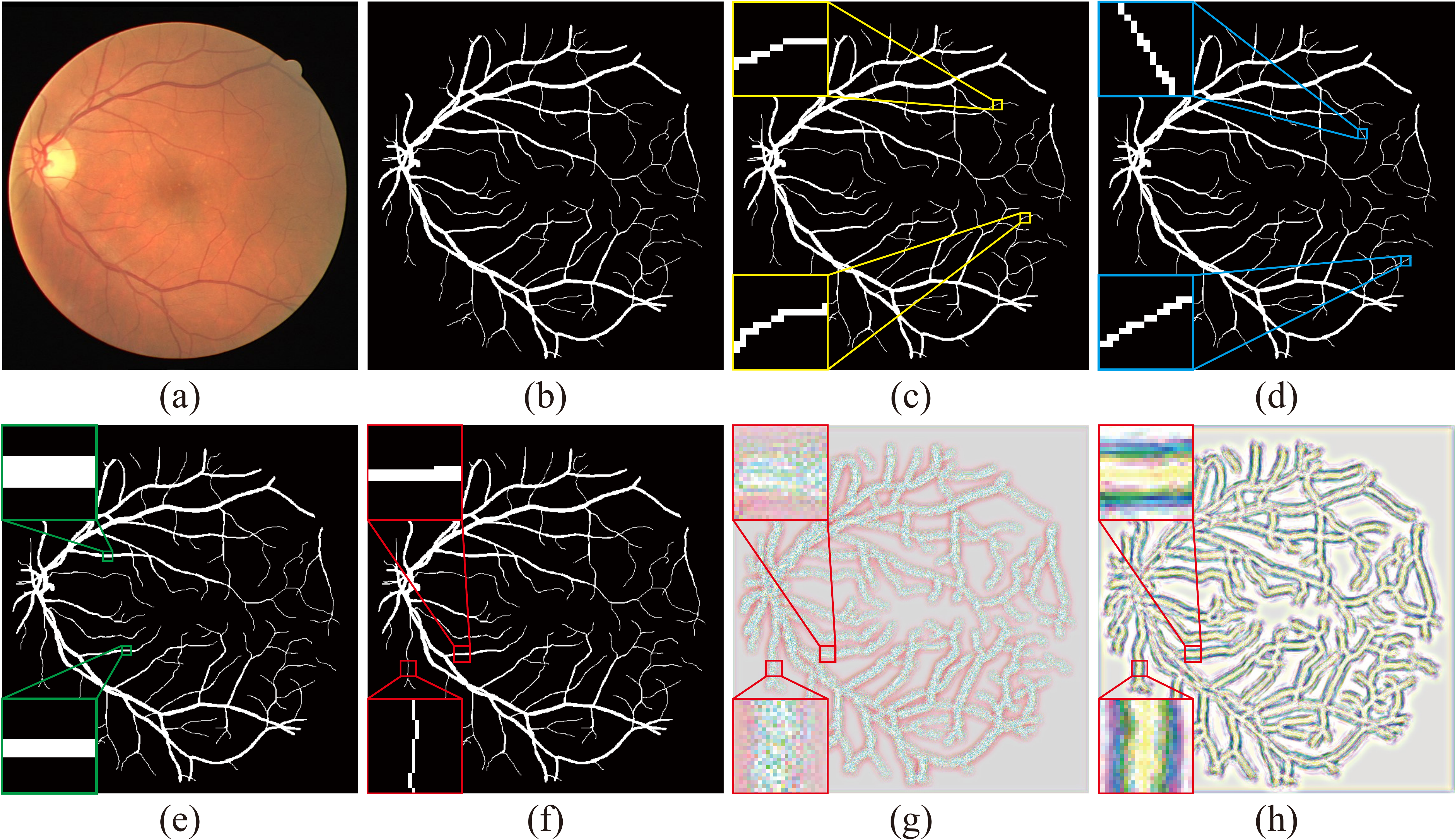}}
\caption{Illustration of the symmetries existing in retinal vessels and the performance of the rotation-and-scale equivariance. (a) A typical retinal image. (b) The corresponding retinal vessel image. (c) Local vascular patterns with the translation symmetry. (d) Local vascular patterns with the rotation symmetry. (e) Local vascular patterns with the scale symmetry. (f) Local vascular patterns with the rotation-and-scale symmetry. (g) Outputs of the randomly initialized traditional convolution filters. (h) Outputs of the randomly initialized RSF-Convs (ours).}
\label{IntroFig}
\end{figure}

\IEEEPARstart{M}orphological changes of the retinal vasculature are substantially relevant to certain diseases, such as diabetes, glaucoma, and hypertension~\cite{abramoff2010retinal}.
Automated segmented vessels can help physicians diagnose and quantify vascular abnormalities in an early stage. However, in clinical practice, it is subjective and laborious for professional ophthalmologists to manually annotate all tiny vessels in fundus images.
Therefore, it is of great clinical significance to develop an automatic algorithm for retinal vessel segmentation~\cite{li2018large}.
Yet retinal vessel segmentation is still a challenging task.
The most formidable aspect lies in the great variation of orientations and scales in the complex retinal vessels. It presents a challenge in characterizing the intricate morphology of the retinal vasculature with pixel-level accuracy, which has a significant impact on the ultimate segmentation results~\cite{mo2017multi}.

To characterize the complex vascular structure, traditional image processing methods~\cite{mendonca2006segmentation, chaudhuri1989detection, zana2001segmentation, singh2023features} heavily rely on handcrafted features and domain knowledge to utilize certain priors present in certain retinal datasets. It often results in either unsatisfactory segmentation results due to the over-simplified priors, or overfitting on one dataset with failure to generalize to another, which makes it difficult to apply in actual clinical practice. 

Recently, deep learning has achieved remarkable performance in retinal vessel segmentation, especially the convolutional neural networks (CNNs) based methods~\cite{ronneberger2015u, krizhevsky2017imagenet, li2020iternet, 9535112, 9714302, singh2024deep}.
One of the key advantages that enable CNNs to achieve such great success is the skillful characterization of the translation symmetry in retinal vessels.
As shown in Fig.~\ref{IntroFig}(c), similar local patterns of retinal vessels tend to appear at different positions in one image, and 
there are numerous similar cases in the whole retinal vasculature. It suggests that this kind of translation symmetry is a widespread image prior existing in retinal vessels, which is successfully embedded into CNNs as a specially designed inner mechanism.
By making use of the dynamic window shifting and the weight sharing, CNNs will always give similar responses to similar patterns, even if they are at different positions,
for which we can claim that CNNs have the translation equivariance~\cite{lecun1995convolutional}. 
It is a distinct advantage for retinal vessel segmentation, and indicates the significance of exploring and making use of more universal image priors of retinal vessels.

Actually, there are still multiple other symmetries existing in retinal vessels, besides the translation symmetry. As shown in Fig.~\ref{IntroFig}(d), the similar vascular structures of different orientations imply the existence of the rotation symmetry in retinal vessels, while the similar local patterns of different scales demonstrate the scale symmetry of retinal vessels, as shown in Fig.~\ref{IntroFig}(e). Furthermore, these two symmetries are widely observed in retinal vessels, indicating their status as universal image priors as well. However, regarding the rotation symmetry or the scale symmetry, CNNs lack an inherent mechanism to exploit them in retinal vessel segmentation.

To cope with the issue, data augmentation is one of the
traditional and widely used methods~\cite{barnard1991invariance}. Although it has advantages in handling global symmetric transformations on entire images, data augmentation is not suitable for symmetric changes on local patterns~\cite{sosnovik2019scale}, such as the diversity of vascular orientations and the variety of scales between capillaries and large vessels.
Previous works attempt to pile up convolution filters of different scales as the spatial pyramid to capture multi-scale features, or design constraints to guide filters to learn more symmetries~\cite{gu2019net, cherukuri2019deep, xu2014scale}. 
Compared with data augmentation, these methods have only partially characterized local symmetries, still in a heuristic way. 

In recent years, by sufficiently utilizing symmetric properties of transformation groups, group equivariant convolutional neural networks (G-CNNs)~\cite{cohen2016group} are proposed to inherently embed the rotation symmetry or the scale symmetry to CNNs, respectively. It implies that, even though similar local vessels have different orientations or different scales, G-CNNs will still give similar responses as well, for which we can similarly claim that G-CNNs have the single rotation equivariance or the single scale equivariance, respectively. Therefore, it is reasonable to anticipate that G-CNNs hold immense potential beyond traditional CNNs in the context of retinal vessel segmentation.

However, besides the single rotation symmetry or the single scale symmetry as shown in Figs.~\ref{IntroFig}(d) and (e), it is much more prevalent for retinal vessels to simultaneously have both the rotation symmetry and the scale symmetry. As shown in Fig.~\ref{IntroFig}(f), similar local vascular patterns are rotated to different orientations, while also having different scales. Such cases are so evident and widespread as a more universal image prior existing in retinal vessels, which demonstrates the urgent need to embed not the single rotation symmetry or the single scale symmetry, but the rotation-and-scale symmetry into networks, so that networks will also give similar responses regardless of the orientations and scales of similar local vessels. Similarly, we define it as the rotation-and-scale equivariance. Furthermore, due to the scale variation and the complexity of retinal vessels as shown in Fig.~\ref{IntroFig}(f), there's still a high requirement for the 
 representation accuracy of networks, while maintaining the rotation-and-scale equivariance. With the inherent property of no information loss in both the discrete Fourier transform (DFT) and its inverse (IDFT), the Fourier parameterized scheme exhibits a substantial advantage for accuracy improvement~\cite{xie2022fourier}.

Out of consideration for the aforementioned issues, this study explores a rotation-and-scale equivariant high-accuracy convolution specifically for retinal vessel segmentation. The main contribution of this work can be summarized as follows:

1) We propose a rotation-and-scale equivariant Fourier parameterized convolution (RSF-Conv) 
specifically 
for retinal vessel segmentation. The proposed RSF-Conv successfully 
characterizes the rotation-and-scale symmetry existing in retinal vessels. 
As shown in Figs.~\ref{IntroFig}(g) and (h), by passing a retinal vessel image into random initialized traditional convolution filters and RSF-Convs, the output of traditional convolution filters are chaotic and unstructured, while RSF-Convs tend to characterize the similar retinal vessels of different orientations and scales with similar structured local patterns. It demonstrates that RSF-Conv is equipped with the rotation-and-scale equivariance.


2) RSF-Conv serves as a general module, capable of being seamlessly integrated into existing networks in a plug-and-play manner, while significantly reducing the number of parameters. Specifically, without changing the network architectures, we replace the traditional convolution filters in typical methods, i.e., U-Net~\cite{ronneberger2015u} and Iter-Net~\cite{li2020iternet}, with the proposed RSF-Convs. Both RSF-Conv+U-Net and RSF-Conv+Iter-Net have merely 13.9\% parameters of the corresponding backbones.

3) We faithfully reproduce multiple state-of-the-art methods and conduct comprehensive experiments under identical conditions.
RSF-Conv+U-Net and RSF-Conv+Iter-Net not only have slight advantages under in-domain evaluation, but more importantly, outperform all comparison methods by a significant margin under out-of-domain evaluation. It demonstrates greater practical clinical significance for the prevalent cross-device and cross-hospital challenges in clinical practice. To comprehensively demonstrate the effectiveness of RSF-Conv, we also apply RSF-Conv+U-Net and RSF-Conv+Iter-Net to retinal artery/vein classification and achieve promising performance as well. These illustrate the high pixel-level accuracy, remarkable generalization, and clinical application potential of RSF-Conv.

\section{Related Work}

\subsection{Retinal Vessel Segmentation}

Early studies about retinal vessel segmentation are mostly based on the traditional image processing~\cite{ mendonca2006segmentation, chaudhuri1989detection, zana2001segmentation}, such as handcraft filters and morphological operations, which generally have an unsatisfactory performance.
Recently, deep learning-based methods have achieved remarkable performance, among which U-Net~\cite{ronneberger2015u} is one of the most widely used methods in medical image segmentation. Despite the excellent accuracy compared with the traditional methods, U-Net is still not effective enough to handle the segmentation for complex retinal vasculature.

To alleviate the loss of spatial information caused by consecutive pooling operations in U-Net, CE-Net~\cite{gu2019net} adopted a multi-scale branch structure with dilated dense blocks and residual poolings. U-Net++~\cite{zhou2019unet++} redesigned the skip connections in U-Net, which enables more sufficient information fusion between multiple scales. In CS-Net~\cite{mou2019cs}, spatial attention and channel attention were used to advance the fusion of local and global information, which benefits the capture of the vascular morphology. To improve the segmentation connectivity of retinal vessels, Iter-Net~\cite{li2020iternet} cascades a U-Net with several mini U-Nets and shrinks the number of channels, which makes the network both sufficiently deep and lightweight. SCS-Net~\cite{wu2021scs} proposed a feature aggregation module to adjust the receptive fields adaptively, and replaced the skip connection in U-Net with a feature fusion module, which flexibly fuses the spatial and semantic information, and suppresses the false positives from the background. To alleviate the successive resolution loss, RV-GAN~\cite{kamran2021rv} proposed a novel GAN-based model with dual generators and multi-scale discriminators.
SGL~\cite{zhou2021study} proposed a new group learning strategy, which efficiently alleviated the overfitting problem caused by the insufficiency of training datasets.

Very recently, to reduce the time and the required experience in the network design, Genetic U-Net~\cite{9535112} first applied the evolutionary neural architecture search (NAS) to retinal vessel segmentation and achieved an excellent improvement with a compact network structure. Besides, DE-DCGCN-EE~\cite{9714302} constructed a dynamic-channel graph convolution network with dual encoders and edge enhancement, which alleviates the loss of edge information and utilizes topological relations in feature maps. LIOT~\cite{shi2022local} proposed a novel imaging preprocessing, which is more sensitive to curvilinear structures and invariant to contrast perturbation. For the better connectivity of segmented vessels, FR-UNet~\cite{liu2022full} proposed a dual-threshold iterative algorithm to perform segmentation gradually. Compared with CNNs, SwinUnet~\cite{cao2022swin} constructed a U-shaped prue Transformer-based network to model global and long-range information of medical images. To explore and enhance the correlations between the global and local context for better feature extraction, MISSFormer~\cite{huang2022missformer} proposed a novel position-free Transformer with a redesigned encoder-decoder structure.


In this study,
different from the previous methods,
we do not change network architectures, but merely replace the traditional convolution filters in U-Net and Iter-Net with our proposed RSF-Conv. With basic network architectures and training strategies, RSF-Conv has achieved high pixel-level accuracy and remarkable generalization.

\subsection{Equivarient CNNs}

Compared with multi-layer perceptrons (MLPs)
~\cite{rumelhart1985learning}
, CNNs successfully embed translation equivariance into networks and have achieved great improvement in image processing. It arouses widespread interest in how to equip networks with more equivariance. Data augmentation~\cite{krizhevsky2017imagenet} is the most widely used approach, which enables networks to learn symmetries by enriching datasets with multiple transformations. In retinal vessel segmentation,~\cite{uysal2021exploring} found that, with sufficient data augmentation, the vanilla U-Net still can achieve near state-of-the-art performance. The method is straightforward and embeds networks with global symmetries, but accordingly, it is time-consuming and lacks the characterization of the symmetries in local patterns.

One of the first CNN-based networks that focus on the local scale symmetry is SiCNN~\cite{xu2014scale}. It interpolated convolution kernels in multi-columns to force filters to have identical patterns in different scales. For retinal vessel segmentation, DRIS-GP~\cite{cherukuri2019deep} optimized the convolution filters under designed constraints to learn the rotation-and-scale symmetry in the vessel morphology. This category of methods characterizes local symmetries in a heuristic approach, which more or less limits the expressiveness of equivariance.

A series of recent works attempt to incorporate equivariance into networks by utilizing the symmetry of groups. G-CNN~\cite{cohen2016group} firstly constructed the group equivariant framework, which achieves the equivariance to the discrete $\pi/2$ rotation. HexaConv~\cite{hoogeboom2018hexaconv} further extended the discrete rotation group to $\pi/3$ by changing the image representation into hexagonal lattices. Based on the scale-spaces theory, DSS~\cite{worrall2019deep} constructed the scale equivariant framework under the formulation of semi-groups, and used dilated convolutions to represent multi-scales.
It restricts the method only to integer rescaling factors.
By rearranging the convolution structure, these methods embed the equivariance in an explicit expression. However, bounded by the traditional discrete convolution operation, it is still difficult for the methods to represent groups adequately.

Currently, the filter parameterization strategy is proposed to address the aforementioned problem. By utilizing harmonic functions as bases, SFCNN~\cite{weiler2018learning} and E2-CNN~\cite{weiler2019general} enabled the parameterized convolution filters to arbitrary rotation angles. PDO-eConv~\cite{shen2020pdo} utilized the partial differential operators to impose the rotation equivariance and firstly derived the boundary of the approximation error for the rotation discretization process. With Hermite polynomials as the bases, SESN~\cite{sosnovik2019scale} expanded the factors of scale groups from integers to the continuous domain. The parameterization approach breaks the limitation of the traditional convolution, and makes it possible to transform filters continuously. However, these methods still suffer from problems in the expression accuracy of the parameterized bases, which results in an inferior performance for tasks that require high accuracy. To address the problem, F-Conv~\cite{xie2022fourier} proposed a novel parameterization scheme based on the 2D symmetric Fourier series expansion, which enables the parameterized filters to achieve high accuracy in both static and transformed cases.

Considering the rotation-and-scale symmetry of retinal vessels as shown in Fig.~\ref{IntroFig}(f) and the requirement for high accuracy, we equip the traditional convolution filters with the rotation-and-scale equivariance as shown in Fig.~\ref{IntroFig}(g) and (h), and adopt the enhanced Fourier basis functions of F-Conv as the parameterization scheme to attain a high representation accuracy. Then, we obtain RSF-Conv, which achieves an excellent performance in retinal vessel segmentation.

\section{Preliminary Knowledge}\label{preliminary}

\begin{figure}[!t]
\centerline{\includegraphics[width=\linewidth]{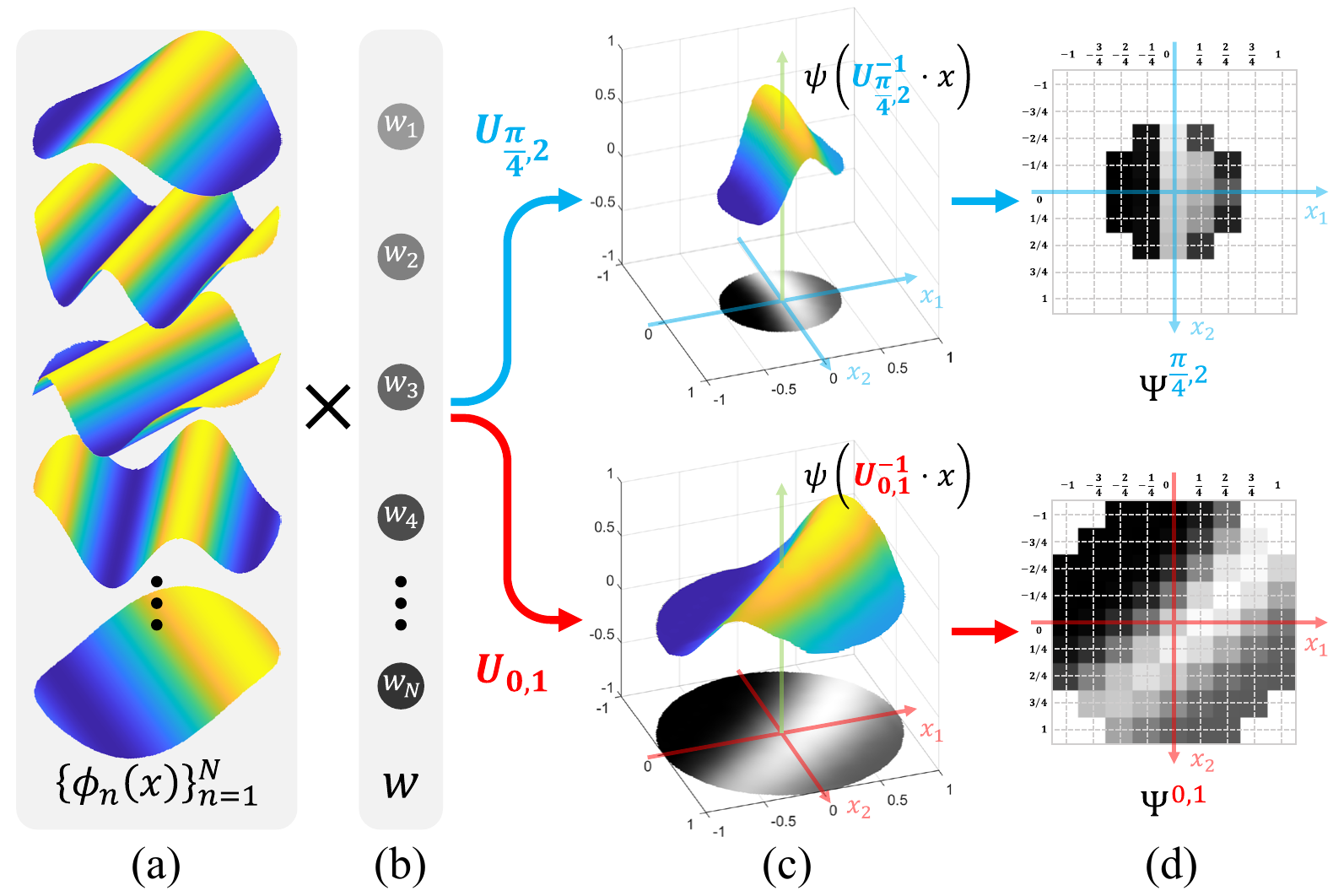}}
\caption{Illustration of the generating filters $\Psi$ at different orientations and scales. (a) The fixed basis functions. (b) The shared coefficient parameters. (c) Latent functions $\psi(x)$ at different orientations and scales. (d) Generated filters at different orientations and scales.}
\label{Parametrization}
\end{figure}

It is not feasible for traditional convolution filters to be rotatable and scalable while keeping learnable, since they are 2D discrete arrays, which are hard to rotate or rescale with a high precision. Thus, it is necessary to represent traditional convolution filters with 2D continuous functions instead of the original 2D discrete arrays, while the latent 2D functions are supposed to be learnable. Formally, as shown in Fig.~\ref{Parametrization}(c) and (d), the elements of a traditional convolution filter ${\Psi}\in\mathbb{R}^{p\times p}$ are represented as:
\begin{equation}\label{Psi_hat}
{\Psi}_{ij} = \psi\l(x_{ij}\r), \quad \forall i,j  = 1,2,\cdots\!, p,
\end{equation}
where $x_{ij}=\left[(i-\frac{p+1}{2})h, (j-\frac{p+1}{2})h\right]^T \in \mathbb{R}^2$ denotes the $p\times p$ mesh grid on 2D spatial coordinates with center as the origin; $p$ and $h$ are the filter size and the mesh size, respectively; and $\psi$ is the latent 2D continuous function (which is learnable). We call $\Psi$  a discretization of $\psi(x)$.

Previous works exploit the linear combination-based filter parameterization technique for representing $\psi(x)$~\cite{weiler2018learning,weiler2019general, shen2020pdo,  xie2022fourier}.
Specifically, as shown in Fig. \ref{Parametrization}(a) and (b), we have
\begin{equation}\label{psi}
  \psi(x)=\sum_{n=1}^N w_n \phi_n(x),
\end{equation}
where $\phi_{n}(x), n = 1,2,\cdots\!, N$ are fixed basis functions, $N$ is the number of fixed basis functions for representing $\psi(x)$, and $w_n$ are learnable coefficients.

As shown in Figs.~\ref{Parametrization}(b), (c), and (d), by adopting coordinate transformation on $\psi(x)$ (or $\phi_n(x)$), we can obtain the rotated and rescaled filter $\Psi^{\theta, s}$, whose elements are defined as:
\begin{equation}\label{Psi_hat2}
\Psi^{\theta, s}_{ij} = \psi\l(U_{\theta,s}^{-1}\cdot x_{ij}\r)=\sum_{n=1}^N w_n \phi_n\l(U_{\theta,s}^{-1}\cdot x_{ij}\r),
\end{equation}
where $x_{ij}$ are points on the mesh grid, $\theta$ is the rotation angle, $s$ is the scale level, and $U_{\theta,s}$ is the rotation-and-scale transformation matrix, i.e.,
\begin{equation}
U_{\theta, s}=s\cdot\left[\begin{array}{cc}
\cos (\theta) & \sin (\theta) \\
-\sin (\theta) & \cos (\theta)
\end{array}\right].
\end{equation}

It is straightforward to see that, as convolution filters in networks, $\Psi^{\theta, s}$ are not only rotatable and scalable with different $\theta$ and $s$, but also learnable with shared parameters $w_n$. We refer to $\psi(x)$ as the parameterized convolution filters.

\section{Methodology}

To characterize the rotation-and-scale symmetry of retinal vessels as shown in Fig.~\ref{IntroFig}(f), we first introduce the proposed rotation-and-scale equivariant Fourier parameterized convolution (RSF-Conv) in the continuous domain and subsequently derive its discretization for concrete implementation.

\subsection{RSF-Conv in Continuous Domain}\label{RS_Framework}

In particular, the initial equivariant convolution for the first layer of the network and the intermediate equivariant convolutions for other layers will be needed.
Similar to the relationship between $\Psi$ and $\psi(x)$ in Eq.~\eqref{Psi_hat}, we model input images as 2D continuous functions $r(x)$,
and feature maps as 2D mappings $f_{(\theta,s)}(x)$ which are indexed by the rotation angle $\theta$ and the scale level $s$, two extra channel dimensions in equivariant convolutions. 

\textbf{Initial Equivariant Convolution.} Specifically, in the continuous domain, the initial equivariant convolution $\Psi^R$ maps the input images $r$ to the feature maps $f$, i.e., $f_{(\theta, s)}(x) = [\Psi^R\circ r]_{(\theta, s)}(x)$. Formally, it is defined as:
\begin{equation} \label{InitConv}
\begin{aligned}\relax
    [\Psi^R\circ r]_{(\theta, s)}(x)=\int_{\mathbb{R}^2} \mu^{-2s} \psi\left(U_{\theta,\mu^s}^{-1}\tilde{x}\right) \cdot r(x + \tilde{x}) d\sigma(\tilde{x}),
\end{aligned}
\end{equation}
where $\mu$ is the step size for scaling the filters and $\sigma$ denotes Haar measure. By substituting convolution filters $\psi(x)$ with the parameterized convolution
filters in Eq.~\eqref{psi}, it is easy to transform convolution filters by rotating and rescaling the fixed basis functions $\phi_{n}(x), n = 1,2,\cdots\!, N$, and perform gradient feedback on the parameters $w_n$.


\textbf{Intermediate Equivariant Convolution.} The intermediate equivariant convolution $\Psi^H$  maps the input feature maps $f$ to the output feature maps $\hat{f}$, i.e., $\hat{f}_{(\theta, s)}(x) = [\Psi^H\circ f]_{(\theta,s)}(x)$. Formally, it is defined as:
\begin{equation} \label{InterConv}
\begin{aligned}\relax
    [\Psi^H\circ f]_{(\theta,s)}(x
    )&=
    \int_{R}\int_{S}\int_{\mathbb{R}^2}\! \mu^{-2s} \psi_{(\tilde{\theta}-\theta, \tilde{s}-\!s)}\!\left(U_{\theta,\mu^s}^{-1}\tilde{x}\right) 
    \\ 
    &\cdot f_{(\tilde{\theta}, \tilde{s})}(x + \tilde{x})d\sigma(\tilde{x})d\sigma(\tilde{s})d\sigma(\tilde{\theta}),
 \end{aligned}
\end{equation}
where $R$ is the rotation transformation group indexed by $\theta$; $S$ is the scale transformation group indexed by $s$;
$\psi_{(\theta,s)}(x)$ defines the parameterized filter with index $\theta$ and $s$, two extra channel dimensions in equivariant convolutions. Similar to Eq.~\eqref{InitConv},  by substituting convolution filters $\psi_{(\theta,s)}(x)$ with the parameterized convolution filters in Eq.~\eqref{psi}, convolution filters are also rotatable and scalable with the fixed basis functions $\phi_{n}(x), n = 1,2,\cdots\!, N$, and learnable with the coefficients $w_n$.

\textbf{Equivariance Analysis.}
Based on the theoretic analysis in previous works~\cite{cohen2016group, xie2022fourier, weiler2018learning, weiler2019general, shen2020pdo}, we have the following results for RSF-Conv in Eqs.~\eqref{InitConv} and~\eqref{InterConv}. 
\begin{customlemma} \label{lemma}
Eqs.~\eqref{InitConv} and~\eqref{InterConv} satisfy the  following equations:
\begin{equation}
\begin{gathered}
\Psi^{R}\circ \pi^{R}_{\hat{\theta},\hat{s}}[r] = \pi_{\hat{\theta},\hat{s}}^{H}[\Psi^{R}\circ r], 
\\
\Psi^{H}\circ\pi^{H}_{\hat{\theta},\hat{s}}[f] = \pi_{\hat{\theta},\hat{s}}^{H}[\Psi^{H}\circ f],
\end{gathered}
\end{equation}
where $\pi_{\hat{\theta},\hat{s}}^R$ and $\pi_{\hat{\theta},\hat{s}}^H$ are the  transformations\footnote{$\pi^R_{\hat{\theta},\hat{s}}[r](x) \!=\! r\l(U^{-1}_{\hat{\theta},\mu^{\hat{s}}}x\r)$, \
$\pi^H_{\hat{\theta},\hat{s}}[f]_{(\theta,s)}(x) \!=\! f_{(\theta-\hat{\theta},s-\!\hat{s})}\l(U^{-1}_{\hat{\theta},\mu^{\hat{s}}}x\r).$ } (i.e., group elements in R and S,  indexed by $\hat{\theta}$ and $\hat{s}$) acting on input images and feature maps, respectively.
\end{customlemma}

It implies that the convolution filters constructed under Eqs.~\eqref{InitConv} and~\eqref{InterConv} are equivariant with respect to the group elements $\hat{\theta}$ and $\hat{s}$. In other words, for similar local retinal vessels of different orientations and scales as shown in Fig.~\ref{IntroFig}(f), the proposed RSF-Conv will always give similar responses as shown in Fig.~\ref{IntroFig}(h). The proof of Theorem is presented in Appendix for completeness.

\subsection{RSF-Conv in Discrete Domain}

\begin{figure*}[!t]
\centerline{\includegraphics[width=\textwidth]{LaTeX/Method_2nd.pdf}}
\caption{Illustration of RSF-Conv, compared with the traditional convolution. (a)(b) The initial and intermediate equivariant convolution filters of RSF-Conv, with three input and nine output (i.e., 3 in/9 out) channels and 9 in/9 out channels, respectively. The kernels highlighted with the same colored box share the same weights. The discrete rotation group $R$ is \{2$\pi$i/3, i=1, 2, 3\}. The arrow orientations indicate the rotation angles of filters. The discrete scale group $S$ has three scale levels, which are indexed by the number of arrow shafts. The RSF-Conv filters with the same pattern are implemented by cyclically shifting along two dimensions, i.e., the rotation angles and the scale levels. (c)(d) The corresponding traditional convolution filters for comparison, which have no pre-designed structures.}
\label{MethodFig}
\end{figure*}

When applying RSF-Conv to digital images, the discrete versions of Eqs.~\eqref{InitConv} and~\eqref{InterConv} are necessary. In this section, we will provide the formal definitions of the discrete RSF-Conv.

We model input images as 2D continuous functions $r(x)$, whose discretization is the 2D array type of images $I \in \mathbb{R}^{t \times t}$, i.e., for $i,j  = 1,2,\cdots\!, t$,
\begin{equation}
{I}_{ij} = r\l(x_{ij}\r).
\end{equation}

Accordingly, we model feature maps as 2D mappings $f_{(\theta,s)}(x)$. 
The discretization of $f_{(\theta,s)}(x)$ is $F^{\theta, s} \in \mathbb{R}^{t \times t \times m \times n}$, which is a 4D tensor with two spatial dimensions and two extra channel dimensions, i.e., for $i,j  = 1,2,\cdots\!, t$,
\begin{equation}
F^{\theta, s}_{ij} = f_{(\theta,s)}\l(x_{ij}\r),
\end{equation}
where $m$ and $n$ are the number of elements in rotation transformation group $R$ and scale transformation group $S$, respectively.

Similar to the definitions of  $r(x)$ and $f_{(\theta,s)}(x)$, we respectively model the filters of the initial and intermediate equivariant convolution as the 2D latent functions $\psi(x)$ and the indexed functions $\psi_{(\theta,s)}(x)$, whose discretizations are $\tilde{\Psi} \in \mathbb{R}^{p \times p \times m \times n}$ and $\tilde{\Phi} \in \mathbb{R}^{p \times p \times m \times n \times m \times n}$, respectively. For $i,j  = 1,2,\cdots\!, p$, $\theta, \tilde{\theta} \in R$, and $s, \tilde{s} \in S$, we have:
\begin{equation}
\begin{aligned}
\tilde{\Psi}^{\theta, s}_{ij} &= \mu^{-2s}\psi\l(U_{\theta,\mu^s}^{-1} x_{ij}\r), 
\\
\tilde{\Phi}^{\theta, s, \tilde{\theta}, \tilde{s}}_{ij} &= \mu^{-2s}\psi_{(\tilde{\theta},\tilde{s})}\l(U_{\theta,\mu^s}^{-1} x_{ij}\r),    
\end{aligned}
\end{equation}
where $\psi(x)$ and $\psi_{(\theta,s)}(x)$ are the parameterized convolution filters defined in Sec.~\ref{preliminary}.

Consequently, we can define the discrete version of the initial and intermediate equivariant convolution in RSF-Conv (i.e., Eqs.~\eqref{InitConv} and~\eqref{InterConv}) as follows:

\textbf{Initial Equivariant Convolution.} For any $\theta \in R$ and $s \in S$, the discrete version of the initial equivariant convolution in Eq.~\eqref{InitConv} is:
\begin{equation}\label{disInitConv}
\l(\tilde{\Psi} \star I\r)^{\theta, s}=\tilde{\Psi}^{\theta, s} * I,
\end{equation}
where $*$ is the traditional discrete 2D convolution.

\textbf{Intermediate Equivariant Convolution.} For any $\theta \in R$ and $s \in S$, the discrete version of the intermediate equivariant convolution in Eq.~\eqref{InterConv} is:
\begin{equation}\label{disInterConv}
\l(\tilde{\Phi} \star F\r)^{\theta, s}=\sum_{\tilde{\theta} \in R} \sum_{\tilde{s} \in S} \tilde{\Phi}^{\theta, s, \tilde{\theta}-\theta, \tilde{s}-s} * F^{\tilde{\theta}, \tilde{s}}.
\end{equation}


Then, it is obvious that we can implement RSF-Conv via the traditional convolution on $I$ and $F$ with the corresponding pre-designed filters $\tilde{\Psi}$ and $\tilde{\Phi}$. 
For the sake of clarity and readability, we provide the illustration of the discrete RSF-Conv filters in Fig.~\ref{MethodFig} in comparison with the traditional convolution filters. From Figs.~\ref{MethodFig}(a) and (b), we can observe that, in RSF-Conv, a series of convolution filers of different orientations and scales share the same pattern, while the filters of the traditional convolution are independent of each other. Besides, the arrangement of filters in RSF-Conv is carefully designed, e.g., from Fig.~\ref{MethodFig}(b), we can observe that the RSF-Conv filters with the same pattern are implemented by cyclically shifting along the rotation dimension and the scale dimension, which is consistent with Eqs.~\eqref{disInitConv} and~\eqref{disInterConv}.

Meanwhile, due to the complexity of the retinal vascular morphology, there is a high requirement for representation accuracy in retinal vessel segmentation.
Therefore, the choice of the fixed basis functions 
is quite crucial for the discrete equivariant convolution $\tilde{\Psi}$ and $\tilde{\Phi}$, which is highly relevant to the representation capability of parameterized filters~\cite{weiler2018learning, weiler2019general, shen2020pdo, xie2022fourier}. In this work, we adopt the enhanced Fourier basis functions proposed in \cite{xie2022fourier} as the fixed basis functions, which effectively alleviate the aliasing effect arising from the rotation or rescaling of traditional Fourier basis functions, thereby improving the representation accuracy of networks while maintaining equivariance. 
As well as we know, RSF-Conv is the first rotation-and-scale equivariant convolution framework with high precision that is suitable for retinal vessel segmentation.

\begin{remark}
RSF-Conv serves as a general module, capable of being seamlessly integrated into existing networks in a plug-and-play manner, replacing the traditional convolution filters while significantly reducing the number of parameters. Specifically, 
we simply replace the traditional convolution filters in two typical methods, U-Net~\cite{ronneberger2015u} and Iter-Net~\cite{li2020iternet}, with our proposed RSF-Conv and keep network architectures.
Both RSF-Conv+U-Net and RSF-Conv+Iter-Net have merely 13.9\% parameters of the corresponding backbones. More details are demonstrated in experiment evaluations.
\end{remark}

\section{Experiments}

\subsection{Datasets and Evaluation Metrics}

For retinal vessel segmentation, we conduct experiments on the following three widely used public datasets: DRIVE~\cite{staal2004ridge}, STARE~\cite{hoover2000locating}, and CHASE\_DB1~\cite{fraz2012ensemble}.
DRIVE consists of 40 fundus images with a resolution of 584$\times$565 pixels, which are divided into 20 training and 20 testing images.
STARE is composed of 20 retinal images with a size of 700$\times$605 pixels, divided into 16 training and 4 testing images. CHASE\_DB1 contains 28 images with a size of 999$\times$960 pixels, which are split into 20 for training and 8 for testing.
All datasets have two expert annotations and only the first annotation is used as the ground truth according to previous works~\cite{wu2021scs, 9714302, 9535112}.
Field of view masks (FOVs) are offered in DRIVE, but not in STARE and CHASE\_DB1. Therefore, we generate the corresponding FOVs following~\cite{liskowski2016segmenting}. In all experiments, we only calculate evaluation metrics inside FOVs for both comparison methods and ours.

For the assessment of segmentation, we choose the following five commonly used evaluation metrics: sensitivity (Se), specificity (Sp), F1-Score (F1), accuracy (Acc), and area under curve (AUC).

\subsection{Comparison Methods}

In order to make a fair comparison, we faithfully reproduce all comparison methods, including U-Net (2015)~\cite{ronneberger2015u}, Iter-Net (2020)~\cite{li2020iternet}, U-Net++ (2019)~\cite{zhou2019unet++}, CE-Net (2019)~\cite{gu2019net}, CS-Net (2019)~\cite{mou2019cs}, SCS-Net (2021)~\cite{wu2021scs}, Genetic U-Net (2022)~\cite{9535112}, DE-DCGCN-EE (2022)~\cite{9714302}, LIOT (2022)~\cite{shi2022local}, FR-UNet (2022)~\cite{liu2022full}, SwinUnet (2022)~\cite{cao2022swin}, and MISSFormer (2023)~\cite{huang2022missformer}. Especially, besides the original LIOT using Iter-Net as the backbone, we also implement a U-Net version and these two are named "LIOT U-Net" and "LIOT Iter-Net" respectively for distinction. For SwinUnet, we also reproduce two versions, one with the officially provided pre-trained model and the other without, which are named "SwinUnet w" and "SwinUnet w/o" respectively. To validate the effectiveness of RSF-Conv, we replace the traditional convolution filters in two typical methods, U-Net and Iter-Net, with RSF-Convs. Without changing the network architectures, we obtain RSF-Conv+U-Net and RSF-Conv+Iter-Net.

\subsection{Implementation Details}
\label{Implemment details}

We perform all experiments on the PyTorch framework with an NVIDIA 3090 GPU. The Adam optimizer~\cite{kingma2014adam} is used with a learning rate of 0.0002.
For the hyper-parameters of RSF-Conv, we adopt $p$ as $6$, $h$ as $0.5$,
the discrete rotation group $R$ as $\{\frac{i\pi}{4}, i = 0,1, \cdots, 7\}$, and the discrete scale group $S$ as $\{(\frac{5}{4})^i, i = 0,\cdots\!,3\}$, which means $\mu = 1.25$. For the loss function, we only apply the binary cross-entropy loss.
\cite{uysal2021exploring} found that, with sufficient data augmentation, a vanilla U-Net can still achieve near state-of-the-art performance in retinal vessel segmentation. Therefore, for all methods, we sufficiently apply various random data augmentations during training, including rotation, rescaling, flip, shearing, brightness, saturation, and contrast. We randomly extract patches of 256$\times$256 pixels from images with a batch size of 2 to train the network for 200 epochs. During testing, overlapping patches of 256$\times$256 pixels are extracted with the stride of 128, which alleviates border effects. The final segmentation results are obtained by binarizing the predicted probability maps with the threshold of 0.5.
Considering fairness, all experiments are conducted under the same experiment conditions for all methods.

\subsection{In-Domain Evaluation}

\begin{table*}[!t]
\renewcommand\arraystretch{0.6}
\centering
\caption{Numerical results under in-domain evaluation. The best and second best are highlighted in \textbf{bold} and {\ul underline}.}
\label{in-domain data}
\resizebox{\textwidth}{!}{
\begin{tabular}{c|ccccc|ccccc|ccccc}
\hline\hline
\multirow{6}{*}{Method}        & \multicolumn{5}{c|}{\multirow{3}{*}{DRIVE
  $\Rightarrow$   DRIVE}}                                                                                                                                                                                                                                                                                                              & \multicolumn{5}{c|}{\multirow{3}{*}{STARE
  $\Rightarrow$   STARE}}                                                                                                                                                                                                                                                                                                              & \multicolumn{5}{c}{\multirow{3}{*}{CHASE\_DB1
  $\Rightarrow$   CHASE\_DB1}}                                                                                                                                                                                                                                                                                                          \\
                               & \multicolumn{5}{c|}{}                                                                                                                                                                                                                                                                                                                                    & \multicolumn{5}{c|}{}                                                                                                                                                                                                                                                                                                                                    & \multicolumn{5}{c}{}                                                                                                                                                                                                                                                                                                                                     \\
                               & \multicolumn{5}{c|}{}                                                                                                                                                                                                                                                                                                                                    & \multicolumn{5}{c|}{}                                                                                                                                                                                                                                                                                                                                    & \multicolumn{5}{c}{}                                                                                                                                                                                                                                                                                                                                     \\ \cline{2-16} 
\multirow{3}{*}{}              & \multirow{3}{*}{\begin{tabular}[c]{@{}c@{}}Se\\ (\%)\end{tabular}} & \multirow{3}{*}{\begin{tabular}[c]{@{}c@{}}Sp\\ (\%)\end{tabular}} & \multirow{3}{*}{\begin{tabular}[c]{@{}c@{}}F1\\ (\%)\end{tabular}} & \multirow{3}{*}{\begin{tabular}[c]{@{}c@{}}Acc\\ (\%)\end{tabular}} & \multirow{3}{*}{\begin{tabular}[c]{@{}c@{}}AUC\\ (\%)\end{tabular}} & \multirow{3}{*}{\begin{tabular}[c]{@{}c@{}}Se\\ (\%)\end{tabular}} & \multirow{3}{*}{\begin{tabular}[c]{@{}c@{}}Sp\\ (\%)\end{tabular}} & \multirow{3}{*}{\begin{tabular}[c]{@{}c@{}}F1\\ (\%)\end{tabular}} & \multirow{3}{*}{\begin{tabular}[c]{@{}c@{}}Acc\\ (\%)\end{tabular}} & \multirow{3}{*}{\begin{tabular}[c]{@{}c@{}}AUC\\ (\%)\end{tabular}} & \multirow{3}{*}{\begin{tabular}[c]{@{}c@{}}Se\\ (\%)\end{tabular}} & \multirow{3}{*}{\begin{tabular}[c]{@{}c@{}}Sp\\ (\%)\end{tabular}} & \multirow{3}{*}{\begin{tabular}[c]{@{}c@{}}F1\\ (\%)\end{tabular}} & \multirow{3}{*}{\begin{tabular}[c]{@{}c@{}}Acc\\ (\%)\end{tabular}} & \multirow{3}{*}{\begin{tabular}[c]{@{}c@{}}AUC\\ (\%)\end{tabular}} \\
                               &                                                                    &                                                                    &                                                                    &                                                                     &                                                                     &                                                                    &                                                                    &                                                                    &                                                                     &                                                                     &                                                                    &                                                                    &                                                                    &                                                                     &                                                                     \\
                               &                                                                    &                                                                    &                                                                    &                                                                     &                                                                     &                                                                    &                                                                    &                                                                    &                                                                     &                                                                     &                                                                    &                                                                    &                                                                    &                                                                     &                                                                     \\ \hline
\multirow{2}{*}{U-Net}         & \multirow{2}{*}{79.05}                                             & \multirow{2}{*}{98.14}                                             & \multirow{2}{*}{82.43}                                             & \multirow{2}{*}{95.71}                                              & \multirow{2}{*}{98.08}                                              & \multirow{2}{*}{77.81}                                             & \multirow{2}{*}{98.88}                                             & \multirow{2}{*}{82.06}                                             & \multirow{2}{*}{97.05}                                              & \multirow{2}{*}{98.90}                                              & \multirow{2}{*}{78.29}                                             & \multirow{2}{*}{{\ul 98.43}}                                    & \multirow{2}{*}{80.69}                                             & \multirow{2}{*}{96.61}                                              & \multirow{2}{*}{98.61}                                              \\
                               &                                                                    &                                                                    &                                                                    &                                                                     &                                                                     &                                                                    &                                                                    &                                                                    &                                                                     &                                                                     &                                                                    &                                                                    &                                                                    &                                                                     &                                                                     \\
\multirow{2}{*}{Iter-Net}      & \multirow{2}{*}{79.15}                                             & \multirow{2}{*}{98.12}                                             & \multirow{2}{*}{82.44}                                             & \multirow{2}{*}{95.71}                                              & \multirow{2}{*}{98.11}                                              & \multirow{2}{*}{77.62}                                             & \multirow{2}{*}{98.92}                                             & \multirow{2}{*}{82.15}                                             & \multirow{2}{*}{97.07}                                              & \multirow{2}{*}{98.92}                                              & \multirow{2}{*}{79.57}                                             & \multirow{2}{*}{98.23}                                             & \multirow{2}{*}{80.63}                                             & \multirow{2}{*}{96.54}                                              & \multirow{2}{*}{98.57}                                              \\
                               &                                                                    &                                                                    &                                                                    &                                                                     &                                                                     &                                                                    &                                                                    &                                                                    &                                                                     &                                                                     &                                                                    &                                                                    &                                                                    &                                                                     &                                                                     \\
\multirow{2}{*}{U-Net++}       & \multirow{2}{*}{79.54}                                             & \multirow{2}{*}{98.08}                                             & \multirow{2}{*}{82.54}                                             & \multirow{2}{*}{{\ul 95.72}}                                     & \multirow{2}{*}{{\ul 98.14}}                                     & \multirow{2}{*}{79.47}                                             & \multirow{2}{*}{98.82}                                             & \multirow{2}{*}{82.81}                                             & \multirow{2}{*}{97.14}                                              & \multirow{2}{*}{98.92}                                              & \multirow{2}{*}{79.64}                                             & \multirow{2}{*}{98.31}                                             & \multirow{2}{*}{{\ul 81.01}}                                    & \multirow{2}{*}{96.62}                                              & \multirow{2}{*}{98.57}                                              \\
                               &                                                                    &                                                                    &                                                                    &                                                                     &                                                                     &                                                                    &                                                                    &                                                                    &                                                                     &                                                                     &                                                                    &                                                                    &                                                                    &                                                                     &                                                                     \\
\multirow{2}{*}{CE-Net}        & \multirow{2}{*}{76.67}                                             & \multirow{2}{*}{98.15}                                             & \multirow{2}{*}{80.99}                                             & \multirow{2}{*}{95.42}                                              & \multirow{2}{*}{97.75}                                              & \multirow{2}{*}{79.85}                                             & \multirow{2}{*}{98.52}                                             & \multirow{2}{*}{81.70}                                             & \multirow{2}{*}{96.90}                                              & \multirow{2}{*}{98.78}                                              & \multirow{2}{*}{77.90}                                             & \multirow{2}{*}{98.27}                                             & \multirow{2}{*}{79.80}                                             & \multirow{2}{*}{96.43}                                              & \multirow{2}{*}{98.30}                                              \\
                               &                                                                    &                                                                    &                                                                    &                                                                     &                                                                     &                                                                    &                                                                    &                                                                    &                                                                     &                                                                     &                                                                    &                                                                    &                                                                    &                                                                     &                                                                     \\
\multirow{2}{*}{CS-Net}        & \multirow{2}{*}{78.13}                                             & \multirow{2}{*}{98.09}                                             & \multirow{2}{*}{81.71}                                             & \multirow{2}{*}{95.55}                                              & \multirow{2}{*}{97.85}                                              & \multirow{2}{*}{77.70}                                             & \multirow{2}{*}{98.77}                                             & \multirow{2}{*}{81.53}                                             & \multirow{2}{*}{96.95}                                              & \multirow{2}{*}{98.76}                                              & \multirow{2}{*}{78.63}                                             & \multirow{2}{*}{98.31}                                             & \multirow{2}{*}{80.39}                                             & \multirow{2}{*}{96.52}                                              & \multirow{2}{*}{98.44}                                              \\
                               &                                                                    &                                                                    &                                                                    &                                                                     &                                                                     &                                                                    &                                                                    &                                                                    &                                                                     &                                                                     &                                                                    &                                                                    &                                                                    &                                                                     &                                                                     \\
\multirow{2}{*}{SCS-Net}       & \multirow{2}{*}{77.79}                                             & \multirow{2}{*}{98.10}                                             & \multirow{2}{*}{81.53}                                             & \multirow{2}{*}{95.51}                                              & \multirow{2}{*}{97.72}                                              & \multirow{2}{*}{76.71}                                             & \multirow{2}{*}{98.84}                                             & \multirow{2}{*}{81.19}                                             & \multirow{2}{*}{96.92}                                              & \multirow{2}{*}{98.48}                                              & \multirow{2}{*}{77.15}                                             & \multirow{2}{*}{98.20}                                             & \multirow{2}{*}{79.04}                                             & \multirow{2}{*}{96.29}                                              & \multirow{2}{*}{98.06}                                              \\
                               &                                                                    &                                                                    &                                                                    &                                                                     &                                                                     &                                                                    &                                                                    &                                                                    &                                                                     &                                                                     &                                                                    &                                                                    &                                                                    &                                                                     &                                                                     \\
\multirow{2}{*}{Genetic U-Net} & \multirow{2}{*}{78.64}                                             & \multirow{2}{*}{\textbf{98.20}}                                    & \multirow{2}{*}{82.34}                                             & \multirow{2}{*}{95.71}                                              & \multirow{2}{*}{98.09}                                              & \multirow{2}{*}{79.94}                                             & \multirow{2}{*}{98.83}                                             & \multirow{2}{*}{83.15}                                             & \multirow{2}{*}{{\ul 97.19}}                                     & \multirow{2}{*}{{\ul 99.07}}                                     & \multirow{2}{*}{79.85}                                             & \multirow{2}{*}{98.25}                                             & \multirow{2}{*}{80.88}                                             & \multirow{2}{*}{96.58}                                              & \multirow{2}{*}{98.51}                                              \\
                               &                                                                    &                                                                    &                                                                    &                                                                     &                                                                     &                                                                    &                                                                    &                                                                    &                                                                     &                                                                     &                                                                    &                                                                    &                                                                    &                                                                     &                                                                     \\
\multirow{2}{*}{DE-DCGCN-EE}   & \multirow{2}{*}{78.40}                                             & \multirow{2}{*}{{\ul 98.16}}                                    & \multirow{2}{*}{82.08}                                             & \multirow{2}{*}{95.64}                                              & \multirow{2}{*}{97.99}                                              & \multirow{2}{*}{73.98}                                             & \multirow{2}{*}{{\ul 98.96}}                                    & \multirow{2}{*}{80.01}                                             & \multirow{2}{*}{96.79}                                              & \multirow{2}{*}{98.57}                                              & \multirow{2}{*}{76.25}                                             & \multirow{2}{*}{98.35}                                             & \multirow{2}{*}{79.09}                                             & \multirow{2}{*}{96.35}                                              & \multirow{2}{*}{98.17}                                              \\
                               &                                                                    &                                                                    &                                                                    &                                                                     &                                                                     &                                                                    &                                                                    &                                                                    &                                                                     &                                                                     &                                                                    &                                                                    &                                                                    &                                                                     &                                                                     \\
\multirow{2}{*}{FR-UNet}       & \multirow{2}{*}{79.21}                                             & \multirow{2}{*}{98.06}                                             & \multirow{2}{*}{82.29}                                             & \multirow{2}{*}{95.66}                                              & \multirow{2}{*}{98.06}                                              & \multirow{2}{*}{75.52}                                             & \multirow{2}{*}{\textbf{98.97}}                                    & \multirow{2}{*}{81.05}                                             & \multirow{2}{*}{96.94}                                              & \multirow{2}{*}{98.84}                                              & \multirow{2}{*}{78.14}                                             & \multirow{2}{*}{98.27}                                             & \multirow{2}{*}{79.94}                                             & \multirow{2}{*}{96.45}                                              & \multirow{2}{*}{98.43}                                              \\
                               &                                                                    &                                                                    &                                                                    &                                                                     &                                                                     &                                                                    &                                                                    &                                                                    &                                                                     &                                                                     &                                                                    &                                                                    &                                                                    &                                                                     &                                                                     \\
\multirow{2}{*}{LIOT U-Net}    & \multirow{2}{*}{78.74}                                             & \multirow{2}{*}{97.85}                                             & \multirow{2}{*}{81.39}                                             & \multirow{2}{*}{95.42}                                              & \multirow{2}{*}{97.59}                                              & \multirow{2}{*}{77.62}                                             & \multirow{2}{*}{98.76}                                             & \multirow{2}{*}{81.41}                                             & \multirow{2}{*}{96.93}                                              & \multirow{2}{*}{98.66}                                              & \multirow{2}{*}{78.57}                                             & \multirow{2}{*}{98.22}                                             & \multirow{2}{*}{80.01}                                             & \multirow{2}{*}{96.44}                                              & \multirow{2}{*}{98.33}                                              \\
                               &                                                                    &                                                                    &                                                                    &                                                                     &                                                                     &                                                                    &                                                                    &                                                                    &                                                                     &                                                                     &                                                                    &                                                                    &                                                                    &                                                                     &                                                                     \\
\multirow{2}{*}{LIOT Iter-Net} & \multirow{2}{*}{77.35}                                             & \multirow{2}{*}{98.12}                                             & \multirow{2}{*}{81.32}                                             & \multirow{2}{*}{95.48}                                              & \multirow{2}{*}{97.73}                                              & \multirow{2}{*}{78.53}                                             & \multirow{2}{*}{98.69}                                             & \multirow{2}{*}{81.68}                                             & \multirow{2}{*}{96.94}                                              & \multirow{2}{*}{98.78}                                              & \multirow{2}{*}{75.66}                                             & \multirow{2}{*}{{\ul 98.43}}                                    & \multirow{2}{*}{79.07}                                             & \multirow{2}{*}{96.37}                                              & \multirow{2}{*}{98.26}                                              \\
                               &                                                                    &                                                                    &                                                                    &                                                                     &                                                                     &                                                                    &                                                                    &                                                                    &                                                                     &                                                                     &                                                                    &                                                                    &                                                                    &                                                                     &                                                                     \\
\multirow{2}{*}{SwinUnet w/o}   & \multirow{2}{*}{78.40}                                             & \multirow{2}{*}{97.82}                                             & \multirow{2}{*}{81.11}                                             & \multirow{2}{*}{95.35}                                              & \multirow{2}{*}{97.37}                                              & \multirow{2}{*}{73.82}                                             & \multirow{2}{*}{98.80}                                             & \multirow{2}{*}{79.18}                                             & \multirow{2}{*}{96.63}                                              & \multirow{2}{*}{98.07}                                              & \multirow{2}{*}{70.99}                                             & \multirow{2}{*}{97.92}                                             & \multirow{2}{*}{74.00}                                             & \multirow{2}{*}{95.48}                                              & \multirow{2}{*}{96.73}                                              \\
                               &                                                                    &                                                                    &                                                                    &                                                                     &                                                                     &                                                                    &                                                                    &                                                                    &                                                                     &                                                                     &                                                                    &                                                                    &                                                                    &                                                                     &                                                                     \\
\multirow{2}{*}{SwinUnet w}    & \multirow{2}{*}{78.34}                                             & \multirow{2}{*}{98.11}                                             & \multirow{2}{*}{81.91}                                             & \multirow{2}{*}{95.59}                                              & \multirow{2}{*}{97.87}                                              & \multirow{2}{*}{79.19}                                             & \multirow{2}{*}{98.74}                                             & \multirow{2}{*}{82.28}                                             & \multirow{2}{*}{97.04}                                              & \multirow{2}{*}{98.81}                                              & \multirow{2}{*}{78.21}                                             & \multirow{2}{*}{98.35}                                             & \multirow{2}{*}{80.32}                                             & \multirow{2}{*}{96.53}                                              & \multirow{2}{*}{98.37}                                              \\
                               &                                                                    &                                                                    &                                                                    &                                                                     &                                                                     &                                                                    &                                                                    &                                                                    &                                                                     &                                                                     &                                                                    &                                                                    &                                                                    &                                                                     &                                                                     \\
\multirow{2}{*}{MISSFormer}    & \multirow{2}{*}{79.58}                                             & \multirow{2}{*}{97.79}                                             & \multirow{2}{*}{81.74}                                             & \multirow{2}{*}{95.48}                                              & \multirow{2}{*}{97.76}                                              & \multirow{2}{*}{75.46}                                             & \multirow{2}{*}{98.93}                                             & \multirow{2}{*}{80.83}                                             & \multirow{2}{*}{96.90}                                              & \multirow{2}{*}{98.60}                                              & \multirow{2}{*}{76.82}                                             & \multirow{2}{*}{\textbf{98.47}}                                    & \multirow{2}{*}{79.96}                                             & \multirow{2}{*}{96.51}                                              & \multirow{2}{*}{98.45}                                              \\
                               &                                                                    &                                                                    &                                                                    &                                                                     &                                                                     &                                                                    &                                                                    &                                                                    &                                                                     &                                                                     &                                                                    &                                                                    &                                                                    &                                                                     &                                                                     \\ \hline
\multirow{2}{*}{RSF-Conv+U-Net}     & \multirow{2}{*}{{\ul 79.95}}                                    & \multirow{2}{*}{98.05}                                             & \multirow{2}{*}{\textbf{82.70}}                                    & \multirow{2}{*}{\textbf{95.74}}                                     & \multirow{2}{*}{\textbf{98.16}}                                     & \multirow{2}{*}{{\ul 80.02}}                                    & \multirow{2}{*}{98.82}                                             & \multirow{2}{*}{{\ul 83.18}}                                    & \multirow{2}{*}{{\ul 97.19}}                                     & \multirow{2}{*}{98.99}                                              & \multirow{2}{*}{{\ul 81.26}}                                    & \multirow{2}{*}{98.26}                                             & \multirow{2}{*}{\textbf{81.79}}                                    & \multirow{2}{*}{\textbf{96.72}}                                     & \multirow{2}{*}{{\ul 98.74}}                                     \\
                               &                                                                    &                                                                    &                                                                    &                                                                     &                                                                     &                                                                    &                                                                    &                                                                    &                                                                     &                                                                     &                                                                    &                                                                    &                                                                    &                                                                     &                                                                     \\
\multirow{2}{*}{RSF-Conv+Iter-Net}  & \multirow{2}{*}{\textbf{79.97}}                                    & \multirow{2}{*}{98.04}                                             & \multirow{2}{*}{{\ul 82.69}}                                    & \multirow{2}{*}{\textbf{95.74}}                                     & \multirow{2}{*}{\textbf{98.16}}                                     & \multirow{2}{*}{\textbf{80.13}}                                    & \multirow{2}{*}{98.93}                                             & \multirow{2}{*}{\textbf{83.71}}                                    & \multirow{2}{*}{\textbf{97.30}}                                     & \multirow{2}{*}{\textbf{99.14}}                                     & \multirow{2}{*}{\textbf{81.55}}                                    & \multirow{2}{*}{98.22}                                             & \multirow{2}{*}{\textbf{81.79}}                                    & \multirow{2}{*}{{\ul 96.71}}                                     & \multirow{2}{*}{\textbf{98.75}}                                     \\
                               &                                                                    &                                                                    &                                                                    &                                                                     &                                                                     &                                                                    &                                                                    &                                                                    &                                                                     &                                                                     &                                                                    &                                                                    &                                                                    &                                                                     &                                                                     \\ \hline\hline
\end{tabular}
}
\end{table*}

\begin{figure*}[!t]
\centerline{\includegraphics[width=\textwidth]{LaTeX/Indomain_3rd.pdf}}
\caption{Some typical segmentation results under in-domain evaluation.}
\label{in-domain fig}
\end{figure*}

In-domain evaluation implies the training and testing sets are from the same dataset. Given the complex morphology of retinal vasculature, it has a high requirement for the accuracy of methods.
We conduct in-domain evaluations on DRIVE, STARE, and CHASE\_DB1 with identical experimental conditions and calculate metrics inside FOVs.
The experiment results are summarized in Table~\ref{in-domain data} and some visualization results are illustrated in Fig.~\ref{in-domain fig}.

As shown in Table~\ref{in-domain data},  our proposed methods RSF-Conv+U-Net and RSF-Conv+Iter-Net obtain the top two best scores in Se, F1, Acc, and AUC. It indicates that RSF-Conv not only achieves overall better performance (F1, Acc, AUC), but also is capable of capturing more vessels (Se). Although the Sp of RSF-Conv+U-Net and RSF-Conv+Iter-Net are slightly inferior to the best, it is negligible compared with the improvements in Se.
As shown in Fig.~\ref{in-domain fig}, RSF-Conv+U-Net and RSF-Conv+Iter-Net have better identification of small blood vessels and better connectivity and smoothness of large vessels.

These demonstrate that our RSF-Conv is better in both numerical and visualized results. Moreover, such evident improvements are achieved simply by replacing the traditional convolution filters of the backbone methods with our proposed RSF-Conv.  We also compare our method with the Transformer-based methods, i.e., SwinUnet and MISSFormer. RSF-Conv+U-Net and RSF-Conv+Iter-Net also outperform SwinUnet which is fine-tuned based on the pre-trained model. It further indicates the effectiveness of RSF-Conv.

Interestingly, it should be noticed that the sufficient data augmentations aforementioned in Sec.~\ref{Implemment details} are applied for all methods, which include the random rotation in $[0^\circ, 360^\circ]$ and the random rescaling from 0.8 to 1.4. The results in Table~\ref{in-domain data} denote that, even with sufficient global data augmentations, especially rotation and rescaling, RSF-Conv+U-Net and RSF-Conv+Iter-Net still outperform the corresponding backbone methods. It strongly verifies the necessity to characterize the local symmetries of retinal vessels, and demonstrates the effectiveness of RSF-Conv to embed the rotation-and-scale equivariance compared with data augmentation.

\subsection{Out-of-Domain Evaluation}


\begin{table*}[!t]
    \renewcommand\arraystretch{0.6}
    \centering
    \caption{Numerical results under out-of-domain evaluation. The best and second best are highlighted in \textbf{bold} and {\ul underline}.}
    \label{out-of-domain data}

\resizebox{\textwidth}{!}{
\begin{tabular}{c|ccccc|ccccc|ccccc}
\hline\hline
\multirow{6}{*}{Method}            & \multicolumn{5}{c|}{\multirow{3}{*}{DRIVE
  $\Rightarrow$   CHASE\_DB1}}                                                                                                                                                                                                                                                                                    & \multicolumn{5}{c|}{\multirow{3}{*}{DRIVE
  $\Rightarrow$   STARE}}                                                                                                                                                                                                                                                                                         & \multicolumn{5}{c}{\multirow{3}{*}{STARE
  $\Rightarrow$   DRIVE}}                                                                                                                                                                                                                                                                                           \\
                                   & \multicolumn{5}{c|}{}                                                                                                                                                                                                                                                                                                                                    & \multicolumn{5}{c|}{}                                                                                                                                                                                                                                                                                                                                    & \multicolumn{5}{c}{}                                                                                                                                                                                                                                                                                                                                     \\
                                   & \multicolumn{5}{c|}{}                                                                                                                                                                                                                                                                                                                                    & \multicolumn{5}{c|}{}                                                                                                                                                                                                                                                                                                                                    & \multicolumn{5}{c}{}                                                                                                                                                                                                                                                                                                                                     \\ \cline{2-16} 
                                   & \multirow{3}{*}{\begin{tabular}[c]{@{}c@{}}Se\\ (\%)\end{tabular}} & \multirow{3}{*}{\begin{tabular}[c]{@{}c@{}}Sp\\ (\%)\end{tabular}} & \multirow{3}{*}{\begin{tabular}[c]{@{}c@{}}F1\\ (\%)\end{tabular}} & \multirow{3}{*}{\begin{tabular}[c]{@{}c@{}}Acc\\ (\%)\end{tabular}} & \multirow{3}{*}{\begin{tabular}[c]{@{}c@{}}AUC\\ (\%)\end{tabular}} & \multirow{3}{*}{\begin{tabular}[c]{@{}c@{}}Se\\ (\%)\end{tabular}} & \multirow{3}{*}{\begin{tabular}[c]{@{}c@{}}Sp\\ (\%)\end{tabular}} & \multirow{3}{*}{\begin{tabular}[c]{@{}c@{}}F1\\ (\%)\end{tabular}} & \multirow{3}{*}{\begin{tabular}[c]{@{}c@{}}Acc\\ (\%)\end{tabular}} & \multirow{3}{*}{\begin{tabular}[c]{@{}c@{}}AUC\\ (\%)\end{tabular}} & \multirow{3}{*}{\begin{tabular}[c]{@{}c@{}}Se\\ (\%)\end{tabular}} & \multirow{3}{*}{\begin{tabular}[c]{@{}c@{}}Sp\\ (\%)\end{tabular}} & \multirow{3}{*}{\begin{tabular}[c]{@{}c@{}}F1\\ (\%)\end{tabular}} & \multirow{3}{*}{\begin{tabular}[c]{@{}c@{}}Acc\\ (\%)\end{tabular}} & \multirow{3}{*}{\begin{tabular}[c]{@{}c@{}}AUC\\ (\%)\end{tabular}} \\
                                   &                                                                    &                                                                    &                                                                    &                                                                     &                                                                     &                                                                    &                                                                    &                                                                    &                                                                     &                                                                     &                                                                    &                                                                    &                                                                    &                                                                     &                                                                     \\
                                   &                                                                    &                                                                    &                                                                    &                                                                     &                                                                     &                                                                    &                                                                    &                                                                    &                                                                     &                                                                     &                                                                    &                                                                    &                                                                    &                                                                     &                                                                     \\ \hline
\multirow{2}{*}{U-Net}             & \multirow{2}{*}{60.29}                                             & \multirow{2}{*}{{\ul 98.00}}                                    & \multirow{2}{*}{66.85}                                             & \multirow{2}{*}{94.58}                                              & \multirow{2}{*}{95.64}                                              & \multirow{2}{*}{76.21}                                             & \multirow{2}{*}{98.55}                                             & \multirow{2}{*}{79.59}                                             & \multirow{2}{*}{96.61}                                              & \multirow{2}{*}{97.62}                                              & \multirow{2}{*}{66.76}                                             & \multirow{2}{*}{\textbf{99.05}}                                    & \multirow{2}{*}{77.05}                                             & \multirow{2}{*}{94.94}                                              & \multirow{2}{*}{96.56}                                              \\
                                   &                                                                    &                                                                    &                                                                    &                                                                     &                                                                     &                                                                    &                                                                    &                                                                    &                                                                     &                                                                     &                                                                    &                                                                    &                                                                    &                                                                     &                                                                     \\
\multirow{2}{*}{Iter-Net}          & \multirow{2}{*}{58.57}                                             & \multirow{2}{*}{97.73}                                             & \multirow{2}{*}{64.58}                                             & \multirow{2}{*}{94.18}                                              & \multirow{2}{*}{93.77}                                              & \multirow{2}{*}{74.61}                                             & \multirow{2}{*}{98.71}                                             & \multirow{2}{*}{79.29}                                             & \multirow{2}{*}{96.62}                                              & \multirow{2}{*}{97.53}                                              & \multirow{2}{*}{67.21}                                             & \multirow{2}{*}{98.84}                                             & \multirow{2}{*}{76.75}                                             & \multirow{2}{*}{94.82}                                              & \multirow{2}{*}{96.49}                                              \\
                                   &                                                                    &                                                                    &                                                                    &                                                                     &                                                                     &                                                                    &                                                                    &                                                                    &                                                                     &                                                                     &                                                                    &                                                                    &                                                                    &                                                                     &                                                                     \\
\multirow{2}{*}{U-Net++}           & \multirow{2}{*}{57.40}                                             & \multirow{2}{*}{{\ul 98.00}}                                    & \multirow{2}{*}{64.68}                                             & \multirow{2}{*}{94.32}                                              & \multirow{2}{*}{93.94}                                              & \multirow{2}{*}{76.60}                                             & \multirow{2}{*}{98.63}                                             & \multirow{2}{*}{80.19}                                             & \multirow{2}{*}{96.72}                                              & \multirow{2}{*}{97.90}                                              & \multirow{2}{*}{68.70}                                             & \multirow{2}{*}{98.75}                                             & \multirow{2}{*}{77.52}                                             & \multirow{2}{*}{94.93}                                              & \multirow{2}{*}{96.66}                                              \\
                                   &                                                                    &                                                                    &                                                                    &                                                                     &                                                                     &                                                                    &                                                                    &                                                                    &                                                                     &                                                                     &                                                                    &                                                                    &                                                                    &                                                                     &                                                                     \\
\multirow{2}{*}{CE-Net}            & \multirow{2}{*}{73.23}                                             & \multirow{2}{*}{96.84}                                             & \multirow{2}{*}{71.46}                                             & \multirow{2}{*}{94.70}                                              & \multirow{2}{*}{96.38}                                              & \multirow{2}{*}{76.42}                                             & \multirow{2}{*}{98.47}                                             & \multirow{2}{*}{79.40}                                             & \multirow{2}{*}{96.56}                                              & \multirow{2}{*}{98.03}                                              & \multirow{2}{*}{69.17}                                             & \multirow{2}{*}{98.35}                                             & \multirow{2}{*}{76.65}                                             & \multirow{2}{*}{94.63}                                              & \multirow{2}{*}{95.98}                                              \\
                                   &                                                                    &                                                                    &                                                                    &                                                                     &                                                                     &                                                                    &                                                                    &                                                                    &                                                                     &                                                                     &                                                                    &                                                                    &                                                                    &                                                                     &                                                                     \\
\multirow{2}{*}{CS-Net}            & \multirow{2}{*}{61.40}                                             & \multirow{2}{*}{97.77}                                             & \multirow{2}{*}{66.83}                                             & \multirow{2}{*}{94.48}                                              & \multirow{2}{*}{94.54}                                              & \multirow{2}{*}{73.87}                                             & \multirow{2}{*}{{\ul 98.72}}                                    & \multirow{2}{*}{78.85}                                             & \multirow{2}{*}{96.56}                                              & \multirow{2}{*}{97.50}                                              & \multirow{2}{*}{69.15}                                             & \multirow{2}{*}{98.18}                                             & \multirow{2}{*}{76.13}                                             & \multirow{2}{*}{94.48}                                              & \multirow{2}{*}{96.03}                                              \\
                                   &                                                                    &                                                                    &                                                                    &                                                                     &                                                                     &                                                                    &                                                                    &                                                                    &                                                                     &                                                                     &                                                                    &                                                                    &                                                                    &                                                                     &                                                                     \\
\multirow{2}{*}{SCS-Net}           & \multirow{2}{*}{44.91}                                             & \multirow{2}{*}{\textbf{98.38}}                                    & \multirow{2}{*}{55.73}                                             & \multirow{2}{*}{93.54}                                              & \multirow{2}{*}{92.21}                                              & \multirow{2}{*}{54.64}                                             & \multirow{2}{*}{\textbf{99.20}}                                    & \multirow{2}{*}{67.02}                                             & \multirow{2}{*}{95.34}                                              & \multirow{2}{*}{93.72}                                              & \multirow{2}{*}{68.40}                                             & \multirow{2}{*}{98.62}                                             & \multirow{2}{*}{76.91}                                             & \multirow{2}{*}{94.77}                                              & \multirow{2}{*}{95.99}                                              \\
                                   &                                                                    &                                                                    &                                                                    &                                                                     &                                                                     &                                                                    &                                                                    &                                                                    &                                                                     &                                                                     &                                                                    &                                                                    &                                                                    &                                                                     &                                                                     \\
\multirow{2}{*}{Genetic U-Net}     & \multirow{2}{*}{59.55}                                             & \multirow{2}{*}{97.69}                                             & \multirow{2}{*}{65.17}                                             & \multirow{2}{*}{94.24}                                              & \multirow{2}{*}{95.12}                                              & \multirow{2}{*}{77.18}                                             & \multirow{2}{*}{{\ul 98.72}}                                    & \multirow{2}{*}{80.96}                                             & \multirow{2}{*}{96.85}                                              & \multirow{2}{*}{98.47}                                              & \multirow{2}{*}{67.31}                                             & \multirow{2}{*}{{\ul 98.96}}                                    & \multirow{2}{*}{77.17}                                             & \multirow{2}{*}{94.93}                                              & \multirow{2}{*}{97.02}                                              \\
                                   &                                                                    &                                                                    &                                                                    &                                                                     &                                                                     &                                                                    &                                                                    &                                                                    &                                                                     &                                                                     &                                                                    &                                                                    &                                                                    &                                                                     &                                                                     \\
\multirow{2}{*}{DE-DCGCN-EE}       & \multirow{2}{*}{59.19}                                             & \multirow{2}{*}{97.45}                                             & \multirow{2}{*}{64.06}                                             & \multirow{2}{*}{93.98}                                              & \multirow{2}{*}{94.10}                                              & \multirow{2}{*}{74.98}                                             & \multirow{2}{*}{98.64}                                             & \multirow{2}{*}{79.24}                                             & \multirow{2}{*}{96.59}                                              & \multirow{2}{*}{97.65}                                              & \multirow{2}{*}{63.50}                                             & \multirow{2}{*}{98.75}                                             & \multirow{2}{*}{73.80}                                             & \multirow{2}{*}{94.26}                                              & \multirow{2}{*}{95.98}                                              \\
                                   &                                                                    &                                                                    &                                                                    &                                                                     &                                                                     &                                                                    &                                                                    &                                                                    &                                                                     &                                                                     &                                                                    &                                                                    &                                                                    &                                                                     &                                                                     \\
\multirow{2}{*}{FR-UNet}           & \multirow{2}{*}{65.48}                                             & \multirow{2}{*}{97.37}                                             & \multirow{2}{*}{68.24}                                             & \multirow{2}{*}{94.48}                                              & \multirow{2}{*}{95.61}                                              & \multirow{2}{*}{79.01}                                             & \multirow{2}{*}{98.29}                                             & \multirow{2}{*}{80.19}                                             & \multirow{2}{*}{96.61}                                              & \multirow{2}{*}{97.80}                                              & \multirow{2}{*}{67.11}                                             & \multirow{2}{*}{98.74}                                             & \multirow{2}{*}{76.39}                                             & \multirow{2}{*}{94.72}                                              & \multirow{2}{*}{96.24}                                              \\
                                   &                                                                    &                                                                    &                                                                    &                                                                     &                                                                     &                                                                    &                                                                    &                                                                    &                                                                     &                                                                     &                                                                    &                                                                    &                                                                    &                                                                     &                                                                     \\
\multirow{2}{*}{LIOT U-Net}        & \multirow{2}{*}{59.51}                                             & \multirow{2}{*}{97.29}                                             & \multirow{2}{*}{63.75}                                             & \multirow{2}{*}{93.87}                                              & \multirow{2}{*}{94.40}                                              & \multirow{2}{*}{77.43}                                             & \multirow{2}{*}{98.45}                                             & \multirow{2}{*}{79.94}                                             & \multirow{2}{*}{96.63}                                              & \multirow{2}{*}{98.07}                                              & \multirow{2}{*}{68.23}                                             & \multirow{2}{*}{97.63}                                             & \multirow{2}{*}{73.96}                                             & \multirow{2}{*}{93.89}                                              & \multirow{2}{*}{96.12}                                              \\
                                   &                                                                    &                                                                    &                                                                    &                                                                     &                                                                     &                                                                    &                                                                    &                                                                    &                                                                     &                                                                     &                                                                    &                                                                    &                                                                    &                                                                     &                                                                     \\
\multirow{2}{*}{LIOT Iter-Net}     & \multirow{2}{*}{62.09}                                             & \multirow{2}{*}{96.79}                                             & \multirow{2}{*}{63.92}                                             & \multirow{2}{*}{93.65}                                              & \multirow{2}{*}{94.35}                                              & \multirow{2}{*}{79.93}                                             & \multirow{2}{*}{98.21}                                             & \multirow{2}{*}{80.42}                                             & \multirow{2}{*}{96.62}                                              & \multirow{2}{*}{98.11}                                              & \multirow{2}{*}{63.48}                                             & \multirow{2}{*}{98.36}                                             & \multirow{2}{*}{72.66}                                             & \multirow{2}{*}{93.92}                                              & \multirow{2}{*}{96.26}                                              \\
                                   &                                                                    &                                                                    &                                                                    &                                                                     &                                                                     &                                                                    &                                                                    &                                                                    &                                                                     &                                                                     &                                                                    &                                                                    &                                                                    &                                                                     &                                                                     \\
\multirow{2}{*}{SwinUnet w/o}       & \multirow{2}{*}{67.76}                                             & \multirow{2}{*}{97.02}                                             & \multirow{2}{*}{68.55}                                             & \multirow{2}{*}{94.37}                                              & \multirow{2}{*}{95.27}                                              & \multirow{2}{*}{77.29}                                             & \multirow{2}{*}{98.49}                                             & \multirow{2}{*}{80.00}                                             & \multirow{2}{*}{96.65}                                              & \multirow{2}{*}{98.01}                                              & \multirow{2}{*}{60.93}                                             & \multirow{2}{*}{98.84}                                             & \multirow{2}{*}{72.15}                                             & \multirow{2}{*}{94.01}                                              & \multirow{2}{*}{94.85}                                              \\
                                   &                                                                    &                                                                    &                                                                    &                                                                     &                                                                     &                                                                    &                                                                    &                                                                    &                                                                     &                                                                     &                                                                    &                                                                    &                                                                    &                                                                     &                                                                     \\
\multirow{2}{*}{SwinUnet w}        & \multirow{2}{*}{76.60}                                             & \multirow{2}{*}{96.16}                                             & \multirow{2}{*}{71.19}                                             & \multirow{2}{*}{94.39}                                              & \multirow{2}{*}{95.93}                                              & \multirow{2}{*}{78.01}                                             & \multirow{2}{*}{98.70}                                             & \multirow{2}{*}{81.40}                                             & \multirow{2}{*}{96.91}                                              & \multirow{2}{*}{98.28}                                              & \multirow{2}{*}{69.28}                                             & \multirow{2}{*}{98.53}                                             & \multirow{2}{*}{77.25}                                             & \multirow{2}{*}{94.81}                                              & \multirow{2}{*}{96.14}                                              \\
                                   &                                                                    &                                                                    &                                                                    &                                                                     &                                                                     &                                                                    &                                                                    &                                                                    &                                                                     &                                                                     &                                                                    &                                                                    &                                                                    &                                                                     &                                                                     \\
\multirow{2}{*}{MISSFormer}        & \multirow{2}{*}{61.56}                                             & \multirow{2}{*}{97.79}                                             & \multirow{2}{*}{66.99}                                             & \multirow{2}{*}{94.51}                                              & \multirow{2}{*}{95.62}                                              & \multirow{2}{*}{77.71}                                             & \multirow{2}{*}{98.58}                                             & \multirow{2}{*}{80.67}                                             & \multirow{2}{*}{96.77}                                              & \multirow{2}{*}{98.30}                                              & \multirow{2}{*}{63.74}                                             & \multirow{2}{*}{98.63}                                             & \multirow{2}{*}{73.63}                                             & \multirow{2}{*}{94.19}                                              & \multirow{2}{*}{95.83}                                              \\
                                   &                                                                    &                                                                    &                                                                    &                                                                     &                                                                     &                                                                    &                                                                    &                                                                    &                                                                     &                                                                     &                                                                    &                                                                    &                                                                    &                                                                     &                                                                     \\ \hline
\multirow{2}{*}{RSF-Conv+U-Net}    & \multirow{2}{*}{{\ul 76.79}}                                    & \multirow{2}{*}{96.87}                                             & \multirow{2}{*}{\textbf{73.76}}                                    & \multirow{2}{*}{\textbf{95.05}}                                     & \multirow{2}{*}{{\ul 96.44}}                                     & \multirow{2}{*}{{\ul 81.63}}                                    & \multirow{2}{*}{98.48}                                             & \multirow{2}{*}{{\ul 82.63}}                                    & \multirow{2}{*}{{\ul 97.02}}                                     & \multirow{2}{*}{{\ul 98.66}}                                     & \multirow{2}{*}{\textbf{75.03}}                                    & \multirow{2}{*}{98.09}                                             & \multirow{2}{*}{{\ul 79.76}}                                    & \multirow{2}{*}{{\ul 95.15}}                                     & \multirow{2}{*}{{\ul 97.09}}                                     \\
                                   &                                                                    &                                                                    &                                                                    &                                                                     &                                                                     &                                                                    &                                                                    &                                                                    &                                                                     &                                                                     &                                                                    &                                                                    &                                                                    &                                                                     &                                                                     \\
\multirow{2}{*}{RSF-Conv+Iter-Net} & \multirow{2}{*}{\textbf{76.95}}                                    & \multirow{2}{*}{96.53}                                             & \multirow{2}{*}{{\ul 72.68}}                                    & \multirow{2}{*}{{\ul 94.76}}                                     & \multirow{2}{*}{\textbf{96.76}}                                     & \multirow{2}{*}{\textbf{82.38}}                                    & \multirow{2}{*}{98.43}                                             & \multirow{2}{*}{\textbf{82.85}}                                    & \multirow{2}{*}{\textbf{97.04}}                                     & \multirow{2}{*}{\textbf{98.85}}                                     & \multirow{2}{*}{{\ul 73.62}}                                    & \multirow{2}{*}{98.40}                                             & \multirow{2}{*}{\textbf{79.77}}                                    & \multirow{2}{*}{\textbf{95.25}}                                     & \multirow{2}{*}{\textbf{97.12}}                                     \\
                                   &                                                                    &                                                                    &                                                                    &                                                                     &                                                                     &                                                                    &                                                                    &                                                                    &                                                                     &                                                                     &                                                                    &                                                                    &                                                                    &                                                                     &                                                                     \\ \hline
\multirow{6}{*}{Method}            & \multicolumn{5}{c|}{\multirow{3}{*}{STARE
  $\Rightarrow$   CHASE\_DB1}}                                                                                                                                                                                                                                                                                    & \multicolumn{5}{c|}{\multirow{3}{*}{CHASE\_DB1
  $\Rightarrow$   STARE}}                                                                                                                                                                                                                                                                                     & \multicolumn{5}{c}{\multirow{3}{*}{CHASE\_DB1
  $\Rightarrow$   DRIVE}}                                                                                                                                                                                                                                                                                      \\
                                   & \multicolumn{5}{c|}{}                                                                                                                                                                                                                                                                                                                                    & \multicolumn{5}{c|}{}                                                                                                                                                                                                                                                                                                                                    & \multicolumn{5}{c}{}                                                                                                                                                                                                                                                                                                                                     \\
                                   & \multicolumn{5}{c|}{}                                                                                                                                                                                                                                                                                                                                    & \multicolumn{5}{c|}{}                                                                                                                                                                                                                                                                                                                                    & \multicolumn{5}{c}{}                                                                                                                                                                                                                                                                                                                                     \\ \cline{2-16} 
                                   & \multirow{3}{*}{\begin{tabular}[c]{@{}c@{}}Se\\ (\%)\end{tabular}} & \multirow{3}{*}{\begin{tabular}[c]{@{}c@{}}Sp\\ (\%)\end{tabular}} & \multirow{3}{*}{\begin{tabular}[c]{@{}c@{}}F1\\ (\%)\end{tabular}} & \multirow{3}{*}{\begin{tabular}[c]{@{}c@{}}Acc\\ (\%)\end{tabular}} & \multirow{3}{*}{\begin{tabular}[c]{@{}c@{}}AUC\\ (\%)\end{tabular}} & \multirow{3}{*}{\begin{tabular}[c]{@{}c@{}}Se\\ (\%)\end{tabular}} & \multirow{3}{*}{\begin{tabular}[c]{@{}c@{}}Sp\\ (\%)\end{tabular}} & \multirow{3}{*}{\begin{tabular}[c]{@{}c@{}}F1\\ (\%)\end{tabular}} & \multirow{3}{*}{\begin{tabular}[c]{@{}c@{}}Acc\\ (\%)\end{tabular}} & \multirow{3}{*}{\begin{tabular}[c]{@{}c@{}}AUC\\ (\%)\end{tabular}} & \multirow{3}{*}{\begin{tabular}[c]{@{}c@{}}Se\\ (\%)\end{tabular}} & \multirow{3}{*}{\begin{tabular}[c]{@{}c@{}}Sp\\ (\%)\end{tabular}} & \multirow{3}{*}{\begin{tabular}[c]{@{}c@{}}F1\\ (\%)\end{tabular}} & \multirow{3}{*}{\begin{tabular}[c]{@{}c@{}}Acc\\ (\%)\end{tabular}} & \multirow{3}{*}{\begin{tabular}[c]{@{}c@{}}AUC\\ (\%)\end{tabular}} \\
                                   &                                                                    &                                                                    &                                                                    &                                                                     &                                                                     &                                                                    &                                                                    &                                                                    &                                                                     &                                                                     &                                                                    &                                                                    &                                                                    &                                                                     &                                                                     \\
                                   &                                                                    &                                                                    &                                                                    &                                                                     &                                                                     &                                                                    &                                                                    &                                                                    &                                                                     &                                                                     &                                                                    &                                                                    &                                                                    &                                                                     &                                                                     \\ \hline
\multirow{2}{*}{U-Net}             & \multirow{2}{*}{67.87}                                             & \multirow{2}{*}{97.32}                                             & \multirow{2}{*}{69.70}                                             & \multirow{2}{*}{94.66}                                              & \multirow{2}{*}{95.81}                                              & \multirow{2}{*}{56.04}                                             & \multirow{2}{*}{{\ul 99.61}}                                    & \multirow{2}{*}{70.01}                                             & \multirow{2}{*}{95.84}                                              & \multirow{2}{*}{97.84}                                              & \multirow{2}{*}{61.83}                                             & \multirow{2}{*}{98.67}                                             & \multirow{2}{*}{72.33}                                             & \multirow{2}{*}{93.98}                                              & \multirow{2}{*}{95.81}                                              \\
                                   &                                                                    &                                                                    &                                                                    &                                                                     &                                                                     &                                                                    &                                                                    &                                                                    &                                                                     &                                                                     &                                                                    &                                                                    &                                                                    &                                                                     &                                                                     \\
\multirow{2}{*}{Iter-Net}          & \multirow{2}{*}{58.36}                                             & \multirow{2}{*}{{\ul 98.08}}                                    & \multirow{2}{*}{65.72}                                             & \multirow{2}{*}{94.49}                                              & \multirow{2}{*}{96.03}                                              & \multirow{2}{*}{54.19}                                             & \multirow{2}{*}{99.59}                                             & \multirow{2}{*}{68.35}                                             & \multirow{2}{*}{95.65}                                              & \multirow{2}{*}{97.45}                                              & \multirow{2}{*}{61.67}                                             & \multirow{2}{*}{98.69}                                             & \multirow{2}{*}{72.27}                                             & \multirow{2}{*}{93.97}                                              & \multirow{2}{*}{95.30}                                              \\
                                   &                                                                    &                                                                    &                                                                    &                                                                     &                                                                     &                                                                    &                                                                    &                                                                    &                                                                     &                                                                     &                                                                    &                                                                    &                                                                    &                                                                     &                                                                     \\
\multirow{2}{*}{U-Net++}           & \multirow{2}{*}{60.63}                                             & \multirow{2}{*}{97.72}                                             & \multirow{2}{*}{66.07}                                             & \multirow{2}{*}{94.36}                                              & \multirow{2}{*}{95.59}                                              & \multirow{2}{*}{55.16}                                             & \multirow{2}{*}{99.55}                                             & \multirow{2}{*}{69.01}                                             & \multirow{2}{*}{95.70}                                              & \multirow{2}{*}{97.42}                                              & \multirow{2}{*}{60.35}                                             & \multirow{2}{*}{98.82}                                             & \multirow{2}{*}{71.66}                                             & \multirow{2}{*}{93.92}                                              & \multirow{2}{*}{95.79}                                              \\
                                   &                                                                    &                                                                    &                                                                    &                                                                     &                                                                     &                                                                    &                                                                    &                                                                    &                                                                     &                                                                     &                                                                    &                                                                    &                                                                    &                                                                     &                                                                     \\
\multirow{2}{*}{CE-Net}            & \multirow{2}{*}{69.43}                                             & \multirow{2}{*}{97.14}                                             & \multirow{2}{*}{70.09}                                             & \multirow{2}{*}{94.63}                                              & \multirow{2}{*}{95.95}                                              & \multirow{2}{*}{63.18}                                             & \multirow{2}{*}{99.19}                                             & \multirow{2}{*}{73.59}                                             & \multirow{2}{*}{96.07}                                              & \multirow{2}{*}{97.51}                                              & \multirow{2}{*}{62.89}                                             & \multirow{2}{*}{98.39}                                             & \multirow{2}{*}{72.31}                                             & \multirow{2}{*}{93.87}                                              & \multirow{2}{*}{94.68}                                              \\
                                   &                                                                    &                                                                    &                                                                    &                                                                     &                                                                     &                                                                    &                                                                    &                                                                    &                                                                     &                                                                     &                                                                    &                                                                    &                                                                    &                                                                     &                                                                     \\
\multirow{2}{*}{CS-Net}            & \multirow{2}{*}{68.46}                                             & \multirow{2}{*}{96.99}                                             & \multirow{2}{*}{68.90}                                             & \multirow{2}{*}{94.40}                                              & \multirow{2}{*}{95.50}                                              & \multirow{2}{*}{51.77}                                             & \multirow{2}{*}{99.58}                                             & \multirow{2}{*}{66.31}                                             & \multirow{2}{*}{95.44}                                              & \multirow{2}{*}{96.54}                                              & \multirow{2}{*}{58.26}                                             & \multirow{2}{*}{98.74}                                             & \multirow{2}{*}{69.82}                                             & \multirow{2}{*}{93.59}                                              & \multirow{2}{*}{94.24}                                              \\
                                   &                                                                    &                                                                    &                                                                    &                                                                     &                                                                     &                                                                    &                                                                    &                                                                    &                                                                     &                                                                     &                                                                    &                                                                    &                                                                    &                                                                     &                                                                     \\
\multirow{2}{*}{SCS-Net}           & \multirow{2}{*}{64.24}                                             & \multirow{2}{*}{97.01}                                             & \multirow{2}{*}{66.14}                                             & \multirow{2}{*}{94.04}                                              & \multirow{2}{*}{95.32}                                              & \multirow{2}{*}{50.24}                                             & \multirow{2}{*}{\textbf{99.69}}                                    & \multirow{2}{*}{65.47}                                             & \multirow{2}{*}{95.40}                                              & \multirow{2}{*}{96.70}                                              & \multirow{2}{*}{54.73}                                             & \multirow{2}{*}{\textbf{99.08}}                                    & \multirow{2}{*}{67.97}                                             & \multirow{2}{*}{93.44}                                              & \multirow{2}{*}{93.84}                                              \\
                                   &                                                                    &                                                                    &                                                                    &                                                                     &                                                                     &                                                                    &                                                                    &                                                                    &                                                                     &                                                                     &                                                                    &                                                                    &                                                                    &                                                                     &                                                                     \\
\multirow{2}{*}{Genetic U-Net}     & \multirow{2}{*}{59.20}                                             & \multirow{2}{*}{97.78}                                             & \multirow{2}{*}{65.24}                                             & \multirow{2}{*}{94.29}                                              & \multirow{2}{*}{95.18}                                              & \multirow{2}{*}{70.83}                                             & \multirow{2}{*}{99.04}                                             & \multirow{2}{*}{78.29}                                             & \multirow{2}{*}{96.59}                                              & \multirow{2}{*}{{\ul 98.46}}                                     & \multirow{2}{*}{61.15}                                             & \multirow{2}{*}{98.58}                                             & \multirow{2}{*}{71.58}                                             & \multirow{2}{*}{93.82}                                              & \multirow{2}{*}{95.65}                                              \\
                                   &                                                                    &                                                                    &                                                                    &                                                                     &                                                                     &                                                                    &                                                                    &                                                                    &                                                                     &                                                                     &                                                                    &                                                                    &                                                                    &                                                                     &                                                                     \\
\multirow{2}{*}{DE-DCGCN-EE}       & \multirow{2}{*}{46.97}                                             & \multirow{2}{*}{\textbf{98.09}}                                    & \multirow{2}{*}{56.54}                                             & \multirow{2}{*}{93.46}                                              & \multirow{2}{*}{93.45}                                              & \multirow{2}{*}{62.84}                                             & \multirow{2}{*}{98.92}                                             & \multirow{2}{*}{72.13}                                             & \multirow{2}{*}{95.79}                                              & \multirow{2}{*}{96.25}                                              & \multirow{2}{*}{60.13}                                             & \multirow{2}{*}{98.80}                                             & \multirow{2}{*}{71.42}                                             & \multirow{2}{*}{93.88}                                              & \multirow{2}{*}{94.50}                                              \\
                                   &                                                                    &                                                                    &                                                                    &                                                                     &                                                                     &                                                                    &                                                                    &                                                                    &                                                                     &                                                                     &                                                                    &                                                                    &                                                                    &                                                                     &                                                                     \\
\multirow{2}{*}{FR-UNet}           & \multirow{2}{*}{58.40}                                             & \multirow{2}{*}{98.06}                                             & \multirow{2}{*}{65.65}                                             & \multirow{2}{*}{94.47}                                              & \multirow{2}{*}{96.09}                                              & \multirow{2}{*}{55.66}                                             & \multirow{2}{*}{99.39}                                             & \multirow{2}{*}{68.70}                                             & \multirow{2}{*}{95.60}                                              & \multirow{2}{*}{96.85}                                              & \multirow{2}{*}{58.75}                                             & \multirow{2}{*}{98.72}                                             & \multirow{2}{*}{70.15}                                             & \multirow{2}{*}{93.63}                                              & \multirow{2}{*}{95.16}                                              \\
                                   &                                                                    &                                                                    &                                                                    &                                                                     &                                                                     &                                                                    &                                                                    &                                                                    &                                                                     &                                                                     &                                                                    &                                                                    &                                                                    &                                                                     &                                                                     \\
\multirow{2}{*}{LIOT U-Net}        & \multirow{2}{*}{66.20}                                             & \multirow{2}{*}{96.95}                                             & \multirow{2}{*}{67.25}                                             & \multirow{2}{*}{94.16}                                              & \multirow{2}{*}{94.76}                                              & \multirow{2}{*}{70.23}                                             & \multirow{2}{*}{98.94}                                             & \multirow{2}{*}{77.42}                                             & \multirow{2}{*}{96.45}                                              & \multirow{2}{*}{98.08}                                              & \multirow{2}{*}{54.69}                                             & \multirow{2}{*}{{\ul 99.05}}                                    & \multirow{2}{*}{67.85}                                             & \multirow{2}{*}{93.40}                                              & \multirow{2}{*}{94.89}                                              \\
                                   &                                                                    &                                                                    &                                                                    &                                                                     &                                                                     &                                                                    &                                                                    &                                                                    &                                                                     &                                                                     &                                                                    &                                                                    &                                                                    &                                                                     &                                                                     \\
\multirow{2}{*}{LIOT Iter-Net}     & \multirow{2}{*}{64.32}                                             & \multirow{2}{*}{97.35}                                             & \multirow{2}{*}{67.38}                                             & \multirow{2}{*}{94.36}                                              & \multirow{2}{*}{94.69}                                              & \multirow{2}{*}{69.37}                                             & \multirow{2}{*}{99.01}                                             & \multirow{2}{*}{77.18}                                             & \multirow{2}{*}{96.44}                                              & \multirow{2}{*}{98.29}                                              & \multirow{2}{*}{56.85}                                             & \multirow{2}{*}{98.84}                                             & \multirow{2}{*}{69.00}                                             & \multirow{2}{*}{93.50}                                              & \multirow{2}{*}{94.71}                                              \\
                                   &                                                                    &                                                                    &                                                                    &                                                                     &                                                                     &                                                                    &                                                                    &                                                                    &                                                                     &                                                                     &                                                                    &                                                                    &                                                                    &                                                                     &                                                                     \\
\multirow{2}{*}{SwinUnet w/o}       & \multirow{2}{*}{54.53}                                             & \multirow{2}{*}{97.95}                                             & \multirow{2}{*}{62.28}                                             & \multirow{2}{*}{94.02}                                              & \multirow{2}{*}{94.58}                                              & \multirow{2}{*}{49.14}                                             & \multirow{2}{*}{99.01}                                             & \multirow{2}{*}{61.59}                                             & \multirow{2}{*}{94.68}                                              & \multirow{2}{*}{94.79}                                              & \multirow{2}{*}{56.58}                                             & \multirow{2}{*}{98.11}                                             & \multirow{2}{*}{66.75}                                             & \multirow{2}{*}{92.83}                                              & \multirow{2}{*}{92.57}                                              \\
                                   &                                                                    &                                                                    &                                                                    &                                                                     &                                                                     &                                                                    &                                                                    &                                                                    &                                                                     &                                                                     &                                                                    &                                                                    &                                                                    &                                                                     &                                                                     \\
\multirow{2}{*}{SwinUnet w}        & \multirow{2}{*}{64.88}                                             & \multirow{2}{*}{97.51}                                             & \multirow{2}{*}{68.33}                                             & \multirow{2}{*}{94.55}                                              & \multirow{2}{*}{95.84}                                              & \multirow{2}{*}{61.90}                                             & \multirow{2}{*}{99.34}                                             & \multirow{2}{*}{73.34}                                             & \multirow{2}{*}{96.10}                                              & \multirow{2}{*}{97.90}                                              & \multirow{2}{*}{{\ul 63.36}}                                    & \multirow{2}{*}{98.16}                                             & \multirow{2}{*}{72.02}                                             & \multirow{2}{*}{93.73}                                              & \multirow{2}{*}{94.84}                                              \\
                                   &                                                                    &                                                                    &                                                                    &                                                                     &                                                                     &                                                                    &                                                                    &                                                                    &                                                                     &                                                                     &                                                                    &                                                                    &                                                                    &                                                                     &                                                                     \\
\multirow{2}{*}{MISSFormer}        & \multirow{2}{*}{59.50}                                             & \multirow{2}{*}{97.70}                                             & \multirow{2}{*}{65.17}                                             & \multirow{2}{*}{94.24}                                              & \multirow{2}{*}{95.84}                                              & \multirow{2}{*}{67.88}                                             & \multirow{2}{*}{99.03}                                             & \multirow{2}{*}{76.23}                                             & \multirow{2}{*}{96.33}                                              & \multirow{2}{*}{97.85}                                              & \multirow{2}{*}{61.86}                                             & \multirow{2}{*}{98.31}                                             & \multirow{2}{*}{71.34}                                             & \multirow{2}{*}{93.67}                                              & \multirow{2}{*}{94.43}                                              \\
                                   &                                                                    &                                                                    &                                                                    &                                                                     &                                                                     &                                                                    &                                                                    &                                                                    &                                                                     &                                                                     &                                                                    &                                                                    &                                                                    &                                                                     &                                                                     \\ \hline
\multirow{2}{*}{RSF-Conv+U-Net}    & \multirow{2}{*}{\textbf{74.55}}                                    & \multirow{2}{*}{97.19}                                             & \multirow{2}{*}{\textbf{73.55}}                                    & \multirow{2}{*}{\textbf{95.14}}                                     & \multirow{2}{*}{\textbf{96.93}}                                     & \multirow{2}{*}{\textbf{72.73}}                                    & \multirow{2}{*}{99.02}                                             & \multirow{2}{*}{\textbf{79.45}}                                    & \multirow{2}{*}{{\ul 96.74}}                                     & \multirow{2}{*}{97.95}                                              & \multirow{2}{*}{62.06}                                             & \multirow{2}{*}{\textbf{99.08}}                                    & \multirow{2}{*}{{\ul 73.72}}                                    & \multirow{2}{*}{\textbf{94.37}}                                     & \multirow{2}{*}{\textbf{96.39}}                                     \\
                                   &                                                                    &                                                                    &                                                                    &                                                                     &                                                                     &                                                                    &                                                                    &                                                                    &                                                                     &                                                                     &                                                                    &                                                                    &                                                                    &                                                                     &                                                                     \\
\multirow{2}{*}{RSF-Conv+Iter-Net} & \multirow{2}{*}{{\ul 72.66}}                                    & \multirow{2}{*}{97.24}                                             & \multirow{2}{*}{{\ul 72.51}}                                    & \multirow{2}{*}{{\ul 95.01}}                                     & \multirow{2}{*}{{\ul 96.85}}                                     & \multirow{2}{*}{{\ul 71.15}}                                    & \multirow{2}{*}{99.21}                                             & \multirow{2}{*}{{\ul 79.29}}                                    & \multirow{2}{*}{\textbf{96.78}}                                     & \multirow{2}{*}{\textbf{98.58}}                                     & \multirow{2}{*}{\textbf{64.03}}                                    & \multirow{2}{*}{98.76}                                             & \multirow{2}{*}{\textbf{74.23}}                                    & \multirow{2}{*}{{\ul 94.34}}                                     & \multirow{2}{*}{{\ul 95.96}}                                     \\
                                   &                                                                    &                                                                    &                                                                    &                                                                     &                                                                     &                                                                    &                                                                    &                                                                    &                                                                     &                                                                     &                                                                    &                                                                    &                                                                    &                                                                     &                                                                     \\ \hline\hline
\end{tabular}
}
\end{table*}

\begin{figure*}[!t]
\vspace{-0.5cm}
\centerline{\includegraphics[width=0.88\textwidth]{LaTeX/Outdomain_2nd.pdf}}
\caption{Some typical segmentation results under out-of-domain evaluation.}
\label{out-of-domain fig}
\vspace{-0.5cm}

\end{figure*}

Out-of-domain evaluation implies that models are trained on one dataset and tested on another, which aligns more closely with real clinical applications and holds greater clinical significance, since the cross-device and cross-hospital settings are prevalent in clinical practice. Accordingly, it is more challenging to the generalization and robustness of methods, compared with in-domain evaluation.

With the same experiment setting, we assess all methods under out-of-domain evaluation on three datasets inside FOVs. It should be noted that, since different datasets are different in resolutions, we test models directly with the original resolutions of the testing datasets, rather than rescale them to fit the resolutions of the training datasets. It is fairly applied to all methods under out-of-domain evaluation.
The numerical results of the six out-of-domain experiments are listed in Table~\ref{out-of-domain data} and some visualization results are shown in Fig.~\ref{out-of-domain fig}.

It is obvious that RSF-Conv+U-Net and RSF-Conv+Iter-Net still achieve the top two best performances in Se, F1, Acc, and AUC, and outperform all comparison methods by a significant margin, which indicates the remarkable generalization and robustness of RSF-Conv and the promising potential for clinical applications. 
Note that the Sp of RSF-Conv+U-Net and RSF-Conv+Iter-Net are still slightly lower than the best in some cases. Considering the notable improvements in other metrics, especially in Se, the effectiveness of RSF-Conv is evident. As shown in Fig.~\ref{out-of-domain fig}, RSF-Conv+U-Net and RSF-Conv+Iter-Net not only achieve the best performance for the segmentation of capillaries and large vessels, but more importantly, have a substantial advantage in terms of segmenting the overall vascular structure. Similarly, we compare with the Transformer-based methods under out-of-domain evaluation, and RSF-Conv+U-Net and RSF-Conv+Iter-Net continue to exhibit an advantage, even when SwinUnet is fine-tuned based on the official pre-trained model.

Moreover, we have also applied sufficient data augmentations as used under in-domain evaluation, including random rotation and rescaling. Without any changes to network architectures, RSF-Conv+U-Net and RSF-Conv+Iter-Net also outperform the backbone methods by an even more evident margin. It further demonstrates the necessity and effectiveness of RSF-Conv to embed the symmetries of retinal vessels into networks, compared with external enhancement strategies, especially data augmentation.

\subsection{Retinal Artery/Vein Classification}

\begin{table}[!t]
    \renewcommand\arraystretch{0.6}
    \centering
    \caption{Numerical results of retinal artery/vein classification. The best and second best are highlighted in \textbf{bold} and {\ul underline}.}
    \label{AV_Table}

\begin{tabular}{c|ccccc}
\hline\hline
\multirow{6}{*}{Method}            & \multicolumn{5}{c}{\multirow{3}{*}{AV-DRIVE
  $\Rightarrow$   AV-DRIVE}}                                                                                                                                                                                                                                                                                     \\
                                   & \multicolumn{5}{c}{}                                                                                                                                                                                                                                                                                                                                     \\
                                   & \multicolumn{5}{c}{}                                                                                                                                                                                                                                                                                                                                     \\ \cline{2-6} 
\multirow{3}{*}{}                  & \multirow{3}{*}{\begin{tabular}[c]{@{}c@{}}$\text{Se}_{\text{AV}}$\\ (\%)\end{tabular}} & \multirow{3}{*}{\begin{tabular}[c]{@{}c@{}}$\text{Sp}_{\text{AV}}$\\ (\%)\end{tabular}} & \multirow{3}{*}{\begin{tabular}[c]{@{}c@{}}$\text{F1}_{\text{AV}}$\\ (\%)\end{tabular}} & \multirow{3}{*}{\begin{tabular}[c]{@{}c@{}}$\text{Acc}_{\text{AV}}$\\ (\%)\end{tabular}} & \multirow{3}{*}{\begin{tabular}[c]{@{}c@{}}$\text{AUC}_{\text{AV}}$\\ (\%)\end{tabular}} \\
                                   &                                                                    &                                                                    &                                                                    &                                                                     &                                                                     \\
                                   &                                                                    &                                                                    &                                                                    &                                                                     &                                                                     \\ \hline
\multirow{2}{*}{Iter-Net}          & \multirow{2}{*}{90.21}                                             & \multirow{2}{*}{93.65}                                             & \multirow{2}{*}{91.09}                                             & \multirow{2}{*}{92.12}                                              & \multirow{2}{*}{96.86}                                              \\
                                   &                                                                    &                                                                    &                                                                    &                                                                     &                                                                     \\
\multirow{2}{*}{U-Net}             & \multirow{2}{*}{92.07}                                             & \multirow{2}{*}{95.00}                                             & \multirow{2}{*}{92.88}                                             & \multirow{2}{*}{93.69}                                              & \multirow{2}{*}{97.55}                                              \\
                                   &                                                                    &                                                                    &                                                                    &                                                                     &                                                                     \\
\multirow{2}{*}{U-Net++}           & \multirow{2}{*}{91.26}                                             & \multirow{2}{*}{94.94}                                             & \multirow{2}{*}{92.40}                                             & \multirow{2}{*}{93.29}                                              & \multirow{2}{*}{97.32}                                              \\
                                   &                                                                    &                                                                    &                                                                    &                                                                     &                                                                     \\
\multirow{2}{*}{MMF-Net}           & \multirow{2}{*}{93.64}                                             & \multirow{2}{*}{{\ul 95.23}}                                    & \multirow{2}{*}{{\ul 93.85}}                                    & \multirow{2}{*}{{\ul 94.52}}                                     & \multirow{2}{*}{{\ul 98.00}}                                     \\
                                   &                                                                    &                                                                    &                                                                    &                                                                     &                                                                     \\ \hline
\multirow{2}{*}{RSF-Conv+Iter-Net} & \multirow{2}{*}{{\ul 93.95}}                                    & \multirow{2}{*}{94.59}                                             & \multirow{2}{*}{93.65}                                             & \multirow{2}{*}{94.30}                                              & \multirow{2}{*}{97.85}                                              \\
                                   &                                                                    &                                                                    &                                                                    &                                                                     &                                                                     \\
\multirow{2}{*}{RSF-Conv+U-Net}    & \multirow{2}{*}{\textbf{95.26}}                                    & \multirow{2}{*}{\textbf{95.39}}                                    & \multirow{2}{*}{\textbf{94.80}}                                    & \multirow{2}{*}{\textbf{95.33}}                                     & \multirow{2}{*}{\textbf{98.18}}                                     \\
                                   &                                                                    &                                                                    &                                                                    &                                                                     &                                                                     \\ \hline\hline
\end{tabular}
\end{table}


\begin{figure*}[!t]
\centerline{\includegraphics[width=\textwidth]{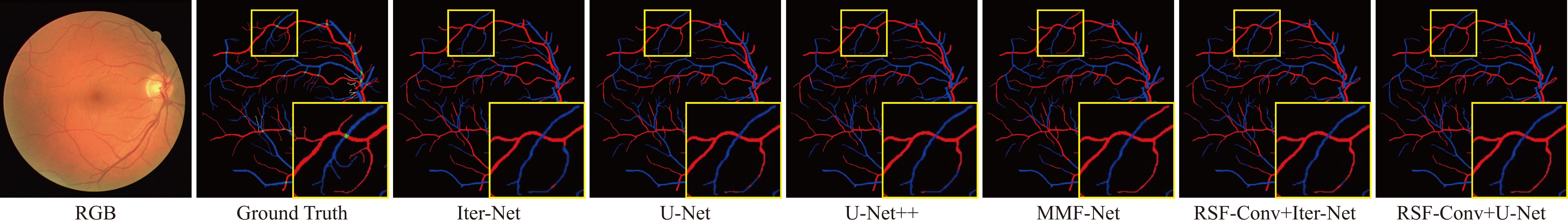}}
\caption{Some typical results of retinal artery/vein classification.}
\label{AV_Fig}
\end{figure*}

To further demonstrate the effectiveness of RSF-Conv, we also apply RSF-Conv+U-Net and RSF-Conv+Iter-Net to retinal artery/vein (A/V) classification, which has a higher requirement for the characterization of retinal vascular morphology. 

We evaluate U-Net~\cite{ronneberger2015u}, Iter-Net~\cite{li2020iternet}, U-Net++~\cite{zhou2019unet++}, RSF-Conv+U-Net, RSF-Conv+Iter-Net and the recent state-of-the-art method in retinal A/V classification, MMF-Net~\cite{yi2023retinal}, using the AV-DRIVE dataset~\cite{hu2013automated}, where arteries, veins, background, and uncertain pixels, are respectively colored red, blue, black, and green within each ground truth.

By viewing arteries as positives and veins as negatives, $\text{sensitivity}_{\text{AV}}$ ($\text{Se}_{\text{AV}}$) and $\text{specificity}_{\text{AV}}$ ($\text{Sp}_{\text{AV}}$) are the metrics to respectively evaluate the capability of detecting arteries and veins, and $\text{F1-Score}_{\text{AV}}$ ($\text{F1}_{\text{AV}}$), $\text{accuracy}_{\text{AV}}$ ($\text{Acc}_{\text{AV}}$) and $\text{area under curve}_{\text{AV}}$ ($\text{AUC}_{\text{AV}}$) indicate the overall performance of methods. Instead of evaluating the A/V classification performance on the segmentation results, we evaluate the performance of all methods on the ground truth artery/vein pixels, which is a more strict criterion, since
the classification of major vessels is a comparatively easier task if capillary vessels are not segmented and thereby excluded from evaluation~\cite{ma2019multi, chen2020tr}. With the sufficient data augmentations aforementioned in Sec.~\ref{Implemment details}, all methods are fairly evaluated under the same experiment settings.

The numerical results are listed in Table~\ref{AV_Table} and some visualization classification results are shown in Fig.~\ref{AV_Fig}. As shown in Fig.~\ref{AV_Fig}, it is much more difficult to distinguish arteries and veins than to distinguish retinal vessels and background. In particular, arteries and veins share almost the same distribution of orientations and scales, which poses a more intricate challenge for methods in characterizing the morphology of retinal vessels.
However, by merely substituting the traditional convolution filters with the proposed RSF-Convs, RSF-Conv+U-Net and RSF-Conv+Iter-Net respectively exhibit a significant improvement over U-Net and Iter-Net. Moreover, RSF-Conv+U-Net outperforms all comparison methods in all metrics, in accordance with the visualization results in Fig.~\ref{AV_Fig}. It further proves the effectiveness of RSF-Conv to characterize the vascular morphology and the great potential for clinical applications.


\subsection{Ablation Study}

In order to verify the effectiveness of the Fourier parameterization, the scale equivariance and the rotation equivariance in RSF-Conv,
we construct the Fourier parameterized convolution (F-Conv), the scale equivariant Fourier parameterized convolution (SF-Conv), and the rotation equivariant Fourier parameterized convolution (RF-Conv) by adjusting the hyper-parameters of the discrete rotation group $R$ and the discrete scale group $S$. Specifically, RF-Conv has only one element in $S$ and SF-Conv has only one element in $R$, while F-Conv respectively has a single element in both $R$ and $S$. We respectively replace the traditional convolution filters in U-Net with the aforementioned variants of RSF-Conv, and obtain F-Conv+U-Net, SF-Conv+U-Net, and RF-Conv+U-Net, as shown in Table~\ref{Ablation data}. 
To exclude other external influencing factors to the ablation study, we do not apply data augmentation to the experiments in this section.

\begin{figure*}[!t]
\centerline{\includegraphics[width=\textwidth]{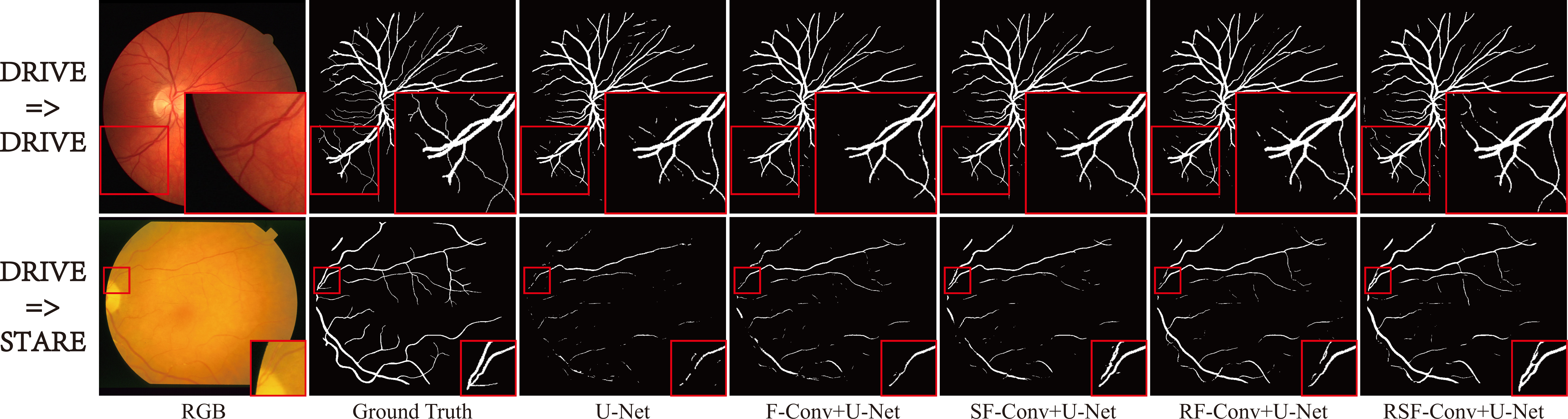}}
\caption{Some typical segmentation results in ablation study.}
\label{Ablation fig}
\end{figure*}

We compare the five networks under two evaluation strategies. The in-domain evaluation is on DRIVE, and the out-of-domain is from DRIVE to STARE. Additionally, we summarize the number of network parameters. The numerical results are listed in Table~\ref{Ablation data}, and some visualization results are shown in Fig.~\ref{Ablation fig}.

\begin{table*}[!t]
    \renewcommand\arraystretch{0.6}
    \centering
    \caption{Ablation studies about RSF-Conv with U-Net as the backbone network. The best is highlighted in \textbf{bold}.}
    \label{Ablation data}
\resizebox{\textwidth}{!}{
\begin{tabular}{c|ccc|ccccc|ccccc}
\hline\hline
\multirow{6}{*}{Method}    & \multirow{6}{*}{\begin{tabular}[c]{@{}c@{}}Rotation\\ Equi.\end{tabular}} & \multirow{6}{*}{\begin{tabular}[c]{@{}c@{}}Scale\\ Equi.\end{tabular}} & \multirow{6}{*}{\begin{tabular}[c]{@{}c@{}}Fourier\\ Param.\end{tabular}} & \multicolumn{5}{c|}{\multirow{3}{*}{DRIVE
  $\Rightarrow$   DRIVE}}                                                                                                                                                                                                                                                                                          & \multicolumn{5}{c}{\multirow{3}{*}{DRIVE
  $\Rightarrow$   STARE}}                                                                                                                                                                                                                                                                                           \\
                           &                                                                           &                                                                        &                                                                           & \multicolumn{5}{c|}{}                                                                                                                                                                                                                                                                                                                                    & \multicolumn{5}{c}{}                                                                                                                                                                                                                                                                                                                                     \\
                           &                                                                           &                                                                        &                                                                           & \multicolumn{5}{c|}{}                                                                                                                                                                                                                                                                                                                                    & \multicolumn{5}{c}{}                                                                                                                                                                                                                                                                                                                                     \\ \cline{5-14} 
                           &                                                                           &                                                                        &                                                                           & \multirow{3}{*}{\begin{tabular}[c]{@{}c@{}}Se\\ (\%)\end{tabular}} & \multirow{3}{*}{\begin{tabular}[c]{@{}c@{}}Sp\\ (\%)\end{tabular}} & \multirow{3}{*}{\begin{tabular}[c]{@{}c@{}}F1\\ (\%)\end{tabular}} & \multirow{3}{*}{\begin{tabular}[c]{@{}c@{}}Acc\\ (\%)\end{tabular}} & \multirow{3}{*}{\begin{tabular}[c]{@{}c@{}}AUC\\ (\%)\end{tabular}} & \multirow{3}{*}{\begin{tabular}[c]{@{}c@{}}Se\\ (\%)\end{tabular}} & \multirow{3}{*}{\begin{tabular}[c]{@{}c@{}}Sp\\ (\%)\end{tabular}} & \multirow{3}{*}{\begin{tabular}[c]{@{}c@{}}F1\\ (\%)\end{tabular}} & \multirow{3}{*}{\begin{tabular}[c]{@{}c@{}}Acc\\ (\%)\end{tabular}} & \multirow{3}{*}{\begin{tabular}[c]{@{}c@{}}AUC\\ (\%)\end{tabular}} \\
                           &                                                                           &                                                                        &                                                                           &                                                                    &                                                                    &                                                                    &                                                                     &                                                                     &                                                                    &                                                                    &                                                                    &                                                                     &                                                                     \\
                           &                                                                           &                                                                        &                                                                           &                                                                    &                                                                    &                                                                    &                                                                     &                                                                     &                                                                    &                                                                    &                                                                    &                                                                     &                                                                     \\ \hline
\multirow{2}{*}{U-Net}     & \multirow{2}{*}{\XSolidBrush}                              & \multirow{2}{*}{\XSolidBrush}                           & \multirow{2}{*}{\XSolidBrush}                              & \multirow{2}{*}{77.35}                                             & \multirow{2}{*}{97.95}                                             & \multirow{2}{*}{80.82}                                             & \multirow{2}{*}{95.33}                                              & \multirow{2}{*}{97.37}                                              & \multirow{2}{*}{51.02}                                             & \multirow{2}{*}{98.93}                                             & \multirow{2}{*}{62.88}                                             & \multirow{2}{*}{94.77}                                              & \multirow{2}{*}{82.69}                                              \\
                           &                                                                           &                                                                        &                                                                           &                                                                    &                                                                    &                                                                    &                                                                     &                                                                     &                                                                    &                                                                    &                                                                    &                                                                     &                                                                     \\
\multirow{2}{*}{F-Conv+U-Net}   & \multirow{2}{*}{\XSolidBrush}                              & \multirow{2}{*}{\XSolidBrush}                           & \multirow{2}{*}{\CheckmarkBold}                            & \multirow{2}{*}{76.59}                                             & \multirow{2}{*}{98.12}                                             & \multirow{2}{*}{80.85}                                             & \multirow{2}{*}{95.38}                                              & \multirow{2}{*}{97.33}                                              & \multirow{2}{*}{58.83}                                             & \multirow{2}{*}{\textbf{98.99}}                                    & \multirow{2}{*}{69.42}                                             & \multirow{2}{*}{95.50}                                              & \multirow{2}{*}{90.20}                                              \\
                           &                                                                           &                                                                        &                                                                           &                                                                    &                                                                    &                                                                    &                                                                     &                                                                     &                                                                    &                                                                    &                                                                    &                                                                     &                                                                     \\
\multirow{2}{*}{SF-Conv+U-Net}  & \multirow{2}{*}{\XSolidBrush}                              & \multirow{2}{*}{\CheckmarkBold}                         & \multirow{2}{*}{\CheckmarkBold}                            & \multirow{2}{*}{76.24}                                             & \multirow{2}{*}{\textbf{98.26}}                                    & \multirow{2}{*}{81.05}                                             & \multirow{2}{*}{95.46}                                              & \multirow{2}{*}{97.65}                                              & \multirow{2}{*}{61.53}                                             & \multirow{2}{*}{98.71}                                             & \multirow{2}{*}{70.28}                                             & \multirow{2}{*}{95.49}                                              & \multirow{2}{*}{95.08}                                              \\
                           &                                                                           &                                                                        &                                                                           &                                                                    &                                                                    &                                                                    &                                                                     &                                                                     &                                                                    &                                                                    &                                                                    &                                                                     &                                                                     \\
\multirow{2}{*}{RF-Conv+U-Net}  & \multirow{2}{*}{\CheckmarkBold}                            & \multirow{2}{*}{\XSolidBrush}                           & \multirow{2}{*}{\CheckmarkBold}                            & \multirow{2}{*}{76.69}                                             & \multirow{2}{*}{98.17}                                             & \multirow{2}{*}{81.05}                                             & \multirow{2}{*}{95.44}                                              & \multirow{2}{*}{97.63}                                              & \multirow{2}{*}{64.81}                                             & \multirow{2}{*}{98.94}                                             & \multirow{2}{*}{73.66}                                             & \multirow{2}{*}{95.98}                                              & \multirow{2}{*}{89.01}                                              \\
                           &                                                                           &                                                                        &                                                                           &                                                                    &                                                                    &                                                                    &                                                                     &                                                                     &                                                                    &                                                                    &                                                                    &                                                                     &                                                                     \\
\multirow{2}{*}{RSF-Conv+U-Net} & \multirow{2}{*}{\CheckmarkBold}                            & \multirow{2}{*}{\CheckmarkBold}                         & \multirow{2}{*}{\CheckmarkBold}                            & \multirow{2}{*}{\textbf{81.18}}                                    & \multirow{2}{*}{97.66}                                             & \multirow{2}{*}{\textbf{82.31}}                                    & \multirow{2}{*}{\textbf{95.56}}                                     & \multirow{2}{*}{\textbf{97.89}}                                     & \multirow{2}{*}{\textbf{69.29}}                                    & \multirow{2}{*}{98.86}                                             & \multirow{2}{*}{\textbf{76.42}}                                    & \multirow{2}{*}{\textbf{96.29}}                                     & \multirow{2}{*}{\textbf{97.65}}                                     \\
                           &                                                                           &                                                                        &                                                                           &                                                                    &                                                                    &                                                                    &                                                                     &                                                                     &                                                                    &                                                                    &                                                                    &                                                                     &                                                                     \\ \hline\hline
\end{tabular}
}
\end{table*}

\subsubsection{In-Domain Ablation}
For in-domain ablation, as shown in Table~\ref{Ablation data}, F-Conv+U-Net performs very close to U-Net. It implies that the Fourier parameterization scheme can approximate the traditional convolution filters with a high accuracy. SF-Conv+U-Net and RF-Conv+U-Net have almost identical performances, both of which are superior to U-Net and F-Conv+U-Net but inferior to RSF-Conv+U-Net. It presents the existence of rotation symmetry and scale symmetry in retinal vessels, and more significantly, demonstrates the necessity to simultaneously embed the rotation-and-scale equivariance into networks. The visual effects of the top row in Fig.~\ref{Ablation fig} are consistent with the numerical results of in-domain ablation in Table~\ref{Ablation data}.

\subsubsection{Out-of-Domain Ablation}
For out-of-domain ablation, as shown in Table~\ref{Ablation data}, F-Conv+U-Net has an overall improvement compared with U-Net. It denotes the stronger generalization capability of the Fourier parameterized convolution filters. SF-Conv+U-Net is superior to RF-Conv+U-Net in AUC but inferior in F1, Acc, Se, and Sp, both of which are still better than U-Net and F-Conv+U-Net, but worse than RSF-Conv+U-Net. It demonstrates the powerful generalization of RSF-Conv and the potential for clinical applications. The visual effects depicted in the bottom row of Fig.~\ref{Ablation fig} align with the numerical results of out-of-domain ablation presented in Table~\ref{Ablation data}.

\subsubsection{Comparison of Parameters}
As aforementioned in Sec.~\ref{Implemment details}, we choose four scale levels as the scale group and eight rotation angles as the rotation group. Since the weights are shared between channels as shown in Fig.~\ref{MethodFig}, the number of parameters of SF-Conv+U-Net, RF-Conv+U-Net, and RSF-Conv+U-Net are about 1/4, 1/8, and 1/32 of F-Conv+U-Net, respectively.
Surprisingly, RSF-Conv+U-Net
achieves such a superior performance with merely 13.9\% parameters of U-Net. 
Similarly, the above analysis is also applicable to RSF-Conv+Iter-Net.
It further illustrates the promising potential of RSF-Conv for clinical deployments on edge devices.


\subsection{Quantitative Equivariance Analysis}

\begin{table}[!t]
    \renewcommand\arraystretch{0.6}
    \centering
    \caption{Quantitative equivariance analysis. The best and second best are highlighted in \textbf{bold} and {\ul underline}.}
    \label{EquiError}
    \resizebox{\columnwidth}{!}{
    \begin{tabular}{c|c|cl|cl|cll}
\hline\hline
\multirow{5}{*}{Method}         & \multirow{5}{*}{Augment}                      & \multicolumn{2}{c|}{\multirow{5}{*}{\begin{tabular}[c]{@{}c@{}}Rotation\\ Equi. Error\end{tabular}}} & \multicolumn{2}{c|}{\multirow{5}{*}{\begin{tabular}[c]{@{}c@{}}Scale\\ Equi. Error\end{tabular}}} & \multicolumn{3}{c}{\multirow{5}{*}{\begin{tabular}[c]{@{}c@{}}Rotation \& Scale\\ Equi. Error\end{tabular}}} \\
                                &                                                & \multicolumn{2}{c|}{}                                                                               & \multicolumn{2}{c|}{}                                                                            & \multicolumn{3}{c}{}                                                                                        \\
                                &                                                & \multicolumn{2}{c|}{}                                                                               & \multicolumn{2}{c|}{}                                                                            & \multicolumn{3}{c}{}                                                                                        \\
                                &                                                & \multicolumn{2}{c|}{}                                                                               & \multicolumn{2}{c|}{}                                                                            & \multicolumn{3}{c}{}                                                                                        \\
                                &                                                & \multicolumn{2}{c|}{}                                                                               & \multicolumn{2}{c|}{}                                                                            & \multicolumn{3}{c}{}                                                                                        \\ \hline
\multirow{4}{*}{U-Net}          & \multirow{2}{*}{\XSolidBrush}   & \multicolumn{2}{c|}{\multirow{2}{*}{0.45}}                                                          & \multicolumn{2}{c|}{\multirow{2}{*}{0.28}}                                                       & \multicolumn{3}{c}{\multirow{2}{*}{0.50}}                                                                   \\
                                &                                                & \multicolumn{2}{c|}{}                                                                               & \multicolumn{2}{c|}{}                                                                            & \multicolumn{3}{c}{}                                                                                        \\
                                & \multirow{2}{*}{\CheckmarkBold} & \multicolumn{2}{c|}{\multirow{2}{*}{0.21}}                                                          & \multicolumn{2}{c|}{\multirow{2}{*}{0.19}}                                                       & \multicolumn{3}{c}{\multirow{2}{*}{0.28}}                                                                   \\
                                &                                                & \multicolumn{2}{c|}{}                                                                               & \multicolumn{2}{c|}{}                                                                            & \multicolumn{3}{c}{}                                                                                        \\ \cline{1-1}
\multirow{4}{*}{RSF-Conv+U-Net} & \multirow{2}{*}{\XSolidBrush}   & \multicolumn{2}{c|}{\multirow{2}{*}{{\ul 0.14}}}                                                    & \multicolumn{2}{c|}{\multirow{2}{*}{{\ul 0.13}}}                                                 & \multicolumn{3}{c}{\multirow{2}{*}{{\ul 0.18}}}                                                             \\
                                &                                                & \multicolumn{2}{c|}{}                                                                               & \multicolumn{2}{c|}{}                                                                            & \multicolumn{3}{c}{}                                                                                        \\
                                & \multirow{2}{*}{\CheckmarkBold} & \multicolumn{2}{c|}{\multirow{2}{*}{\textbf{0.08}}}                                                 & \multicolumn{2}{c|}{\multirow{2}{*}{\textbf{0.10}}}                                              & \multicolumn{3}{c}{\multirow{2}{*}{\textbf{0.13}}}                                                          \\
                                &                                                & \multicolumn{2}{c|}{}                                                                               & \multicolumn{2}{c|}{}                                                                            & \multicolumn{3}{c}{}                                                                                        \\ \hline\hline
\end{tabular}
}

\end{table}

As shown in Figs.~\ref{IntroFig}(g) and (h), we have conducted the qualitative analysis of the rotation-and-scale equivariance. To further quantitatively evaluate the equivariance of RSF-Conv, we compare the equivariance error of RSF-Conv+U-Net and U-Net, respectively. Following previous works~\cite{worrall2019deep, sosnovik2019scale, sosnovik2021disco}, equivariance error is defined as follows: 
\begin{equation}
\text { Equi. Error }=
\frac{
\left\|\pi_{\hat{\theta},\hat{s}}^{H}[\Psi\circ r] - \Psi\circ \pi^{R}_{\hat{\theta},\hat{s}}[r]\right\|_2^2
}{
\left\|\pi_{\hat{\theta},\hat{s}}^{H}[\Psi\circ r]\right\|_2^2
}
,
\end{equation}
where $\Psi$ are fully trained networks. Specifically, the formula calculates the relative percentage error between the network outputs subjected to the random transformations, and the outputs of the network whose inputs undergo the same transformations. There are three types of such transformations, denoted as random rotation, random rescaling, and random rotation-and-rescaling, corresponding to three forms of equivariance, the rotation equivariance, the scale equivariance, and the rotation-and-scale equivariance. We respectively calculate the three types of equivariance error for fully trained RSF-Conv+U-Net and U-Net. To further investigate the effectiveness of RSF-Conv and data augmentation (especially rotation and scale augmentations), during the training process, we separately train both RSF-Conv+U-Net and U-Net under two conditions, one with sufficient data augmentation and the other without.

As shown in Table~\ref{EquiError}, in the case of no data augmentation, the equivariance error of RSF-Conv+U-Net is significantly lower than that of U-Net. Even though data augmentation can substantially reduce the equivariance error for U-Net, the equivariance error of "U-Net with sufficient data augmentation" remains higher than that of "RSF-Conv+U-Net without data augmentation". It quantitatively demonstrates the necessity and effectiveness of RSF-Conv in enhancing equivariance, compared with data augmentation. Moreover, there is no conflict between data augmentation and RSF-Conv. Data augmentation can further reduce the equivariance error of RSF-Conv+U-Net, thereby enhancing equivariance.


\subsection{Future Work}

Although RSF-Conv achieves a satisfactory performance, the truncation of the infinite scale group, due to the limited available resources, still severely affects the representation of equivariance, which is highly relevant to the ultimate segmentation performance. In future work, we will focus on alleviating the effects brought by scale truncation to further improve RSF-Conv and make it more effective for clinical applications.

\section{Conclusion}

In this work, we proposed a rotation-and-scale equivariant Fourier parameterized convolution (RSF-Conv) specifically for retinal vessel segmentation, which effectively characterizes the widespread rotation-and-scale symmetry existing in retinal vessels with pixel-level accuracy. As a general module, RSF-Conv can be seamlessly integrated into existing networks in a plug-and-play manner, while significantly reducing the number of parameters. Specifically, without changing the network architectures, we replaced the traditional convolution filters in typical methods, U-Net and Iter-Net, with the proposed RSF-Convs. The numerical and visualized results illustrated that RSF-Conv+U-Net and RSF-Conv+Iter-Net achieved high pixel-level accuracy and remarkable generalization, with merely 13.9\% parameters of corresponding backbones. It demonstrated the promising potential of RSF-Conv for clinical applications. 


\section*{Appendix}

\setcounter{equation}{0}
\renewcommand\theequation{\arabic{equation}}

\begin{notations}

We model input images as 2D continuous functions $r(x)$ and the rotation-and-scale transformations $\pi^R_{\hat{\theta},\hat{s}}$  (i.e., group elements in R and S,  indexed by $\hat{\theta}$ and $\hat{s}$) act on $r(x)$ by:
\begin{equation}\label{initpiAPP}
    \pi^R_{\hat{\theta},\hat{s}}[r](x) = r\left(U^{-1}_{\hat{\theta},\mu^{\hat{s}}}x\right).
\end{equation}
where $U_{\theta,s}$ is the rotation-and-scale transformation matrix, i.e.,
\begin{equation}\label{TmatirxAPP}
  U_{\theta, s}=s\cdot\left[\begin{array}{cc}
\cos (\theta) & \sin (\theta) \\
-\sin (\theta) & \cos (\theta)
\end{array}\right].
\end{equation}

Feature maps are modeled as 2D mappings $f_{(\theta,s)}(x)$, which are indexed by $\theta$ and $s$, and $\pi^H_{\hat{\theta},\hat{s}}$ are the corresponding transformations (i.e., group elements in R and S,  indexed by $\hat{\theta}$ and $\hat{s}$, respectively), which act on $f_{(\theta,s)}(x)$:
\begin{equation}\label{interpiAPP}
    \pi^H_{\hat{\theta},\hat{s}}[f]_{(\theta,s)}(x) = f_{(\theta-\hat{\theta},s-\hat{s})}\l(U^{-1}_{\hat{\theta},\mu^{\hat{s}}}x\r).
\end{equation}

In the continuous domain, $\Psi^R$ is the initial equivariant convolution, which maps the input image $r$ to the feature maps $f$, i.e., $f_{(\theta, s)}(x) = [\Psi^R\circ r]_{(\theta, s)}(x)$. Formally, it is defined as:
\begin{equation} \label{InitConvAPP}
\begin{aligned}\relax
    [\Psi^R\circ r]_{(\theta, s)}(x)=\int_{\mathbb{R}^2} \mu^{-2s} \psi\left(U_{\theta,\mu^s}^{-1}\tilde{x}\right) \cdot r(x + \tilde{x}) d\sigma(\tilde{x}),
\end{aligned}
\end{equation}
where $\mu$ is the step size for scaling the filters; $\sigma$ denotes the Haar measure on $\mathbb{R}^2$; and $\psi(x)$ is the proposed parameterized filter.

$\Psi^H$ is the intermediate equivariant convolution, which maps the input feature maps $f$ to the output feature maps $\hat{f}$, i.e., $\hat{f}_{(\theta, s)}(x) = [\Psi^H\circ f]_{(\theta,s)}(x)$. Formally, it is defined as:
\begin{equation} \label{InterConvAPP}
\begin{aligned}\relax
    [\Psi^H\circ f]_{(\theta,s)}(x
    )&=
    \int_{R}\int_{S}\int_{\mathbb{R}^2}\! \mu^{-2s} \psi_{(\tilde{\theta}-\theta, \tilde{s}-\!s)}\!\left(U_{\theta,\mu^s}^{-1}\tilde{x}\right) 
    \\ 
    &
    \cdot f_{(\tilde{\theta}, \tilde{s})}(x + \tilde{x})d\sigma(\tilde{x})d\sigma(\tilde{s})d\sigma(\tilde{\theta}),
\end{aligned}
\end{equation}
where $R$ is the rotation transformation group indexed by $\theta$; $S$ is the scale transformation group indexed by $s$; and $\psi_{(\theta,s)}(x)$ defines the parameterized filter with index $\theta$ and $s$ in the two extra dimensions.

\end{notations}

\begin{customlemma}
Eqs.~\eqref{InitConvAPP} and~\eqref{InterConvAPP} satisfy following equations:
\begin{equation}
\begin{gathered}
\Psi^{R}\circ \pi^{R}_{\hat{\theta},\hat{s}}[r] = \pi_{\hat{\theta},\hat{s}}^{H}[\Psi^{R}\circ r], \\
\Psi^{H}\circ\pi^{H}_{\hat{\theta},\hat{s}}[f] = \pi_{\hat{\theta},\hat{s}}^{H}[\Psi^{H}\circ f],
\end{gathered}
\end{equation}
where $\Psi^R$, $\Psi^H$, $\pi_{\hat{\theta},\hat{s}}^H$, and $\pi_{\hat{\theta},\hat{s}}^R$ are defined by Eqs.~\eqref{initpiAPP},~\eqref{interpiAPP},~\eqref{InitConvAPP}, and~\eqref{InterConvAPP}, respectively.
\end{customlemma}

\begin{proof}
Based on the frameworks of the previous works~\cite{cohen2016group, weiler2018learning,weiler2019general, shen2020pdo,  xie2022fourier}, we derive the following proof:

(1) For any $x \in \mathbb{R}^2$, $\theta \in R$, and $s \in S$, we obtain:
\begin{equation}
\begin{aligned}
&\l[\Psi^{R}\circ \pi^{R}_{\hat{\theta},\hat{s}}[r]\r]_{(\theta,s)}(x) 
\\
= &\int_{\mathbb{R}^2} \mu^{-2s} \psi\left(U_{\theta,\mu^{s}}^{-1}\tilde{x}\right) \cdot \pi^{R}_{\hat{\theta},\hat{s}}[r](x + \tilde{x}) d\sigma(\tilde{x}) 
\\
= &\int_{\mathbb{R}^2} \mu^{-2s} \psi\left(U_{\theta,\mu^{s}}^{-1}\tilde{x}\right) \cdot r\l(U^{-1}_{\hat{\theta},\mu^{\hat{s}}}(x + \tilde{x})\r) d\sigma(\tilde{x}).
\end{aligned}
\end{equation}

Let $x' = U^{-1}_{\hat{\theta},\mu^{\hat{s}}}\tilde{x}$, since $|\operatorname{det}(U_{\hat{\theta},\mu^{\hat{s}}})|= \mu^{2\hat{s}}$, and we have:

\begin{equation}
\begin{aligned}
&\int_{\mathbb{R}^2} \mu^{-2s} \psi\left(U_{\theta,\mu^{s}}^{-1}\tilde{x}\right) \cdot r\l(U^{-1}_{\hat{\theta},\mu^{\hat{s}}}(x + \tilde{x})\r) d\sigma(\tilde{x}) \\
= &\int_{\mathbb{R}^2} \mu^{-2(s-\hat{s})} \psi\left((U^{-1}_{\hat{\theta},\mu^{\hat{s}}}U_{\theta,\mu^s})^{-1}x'\right) 
\\
& \qquad \qquad \qquad \qquad \qquad \quad 
\cdot r\l(U^{-1}_{\hat{\theta},\mu^{\hat{s}}}x + x'\r) d\sigma(x') \\
= &\int_{\mathbb{R}^2} \mu^{-2(s-\hat{s})} \psi\left(U^{-1}_{\theta-\hat{\theta},\mu^{s-\hat{s}}}x'\right) 
\\
& \qquad \qquad \qquad \qquad \qquad \quad 
 \cdot r\l(U^{-1}_{\hat{\theta},\mu^{\hat{s}}}x + x'\r) d\sigma(x') \\
= & \ [\Psi^{R}\circ r]_{(\theta-\hat{\theta},s-\hat{s})}(U^{-1}_{\hat{\theta},\mu^{\hat{s}}}x)\\
= & \ \pi_{\hat{\theta},\hat{s}}^{H}[\Psi^{R}\circ r]_{(\theta,s)}(x).
\end{aligned}
\end{equation}

This proves that $\Psi^{R}\circ \pi^{R}_{\hat{\theta},\hat{s}}[r] = \pi_{\hat{\theta},\hat{s}}^{H}[\Psi^{R}\circ r]$.

(2) For any $x \in \mathbb{R}^2$, $\theta \in R$, and $s \in S$, we obtain:
\begin{equation}
\begin{aligned}
&\l[\Psi^{H}\circ \pi^{H}_{\hat{\theta},\hat{s}}[f]\r]_{(\theta,s)}(x) 
\\
=&\int_{R}\int_{S}\int_{\mathbb{R}^2}\! \mu^{-2s} \psi_{(\tilde{\theta}-\theta, \tilde{s}-\!s)}\!\left(U_{\theta,\mu^s}^{-1}\tilde{x}\right) 
\\
& \qquad \qquad 
\cdot \pi^{H}_{\hat{\theta},\hat{s}}[f]_{(\tilde{\theta}, \tilde{s})}(x + \tilde{x})\sigma(\tilde{x})\sigma(\tilde{s})\sigma(\tilde{\theta}) \\
=&\int_{R}\int_{S}\int_{\mathbb{R}^2}\! \mu^{-2s} \psi_{(\tilde{\theta}-\theta, \tilde{s}-\!s)}\!\left(U_{\theta,\mu^s}^{-1}\tilde{x}\right) 
\\ 
& \qquad \qquad 
\cdot f_{(\tilde{\theta}-\hat{\theta}, \tilde{s}-\!\hat{s})}\l(U^{-1}_{\hat{\theta},\mu^{\hat{s}}}(x + \tilde{x})\r)\sigma(\tilde{x})\sigma(\tilde{s})\sigma(\tilde{\theta}).
\end{aligned}
\end{equation}

Let $\tilde{\theta}-\hat{\theta} = \theta'$, $\tilde{s}-\hat{s} = s'$, $x' = U^{-1}_{\hat{\theta},\mu^{\hat{s}}}\tilde{x}$, since  $|\operatorname{det}(U_{\hat{\theta},\mu^{\hat{s}}})|= \mu^{2\hat{s}}$, and we have

\begin{equation}
\begin{aligned}
&\int_{R}\int_{S}\int_{\mathbb{R}^2}\! \mu^{-2s} \psi_{(\tilde{\theta}-\theta, \tilde{s}-\!s)}\!\left(U_{\theta,\mu^s}^{-1}\tilde{x}\right) \
\\
& \qquad \qquad 
\cdot f_{(\tilde{\theta}-\hat{\theta}, \tilde{s}-\!\hat{s})}\l(U^{-1}_{\hat{\theta},\mu^{\hat{s}}}(x + \tilde{x})\r)\sigma(\tilde{x})\sigma(\tilde{s})\sigma(\tilde{\theta}) \\
= &\int_{R}\int_{S}\int_{\mathbb{R}^2}\! \mu^{-2(s-\hat{s})} \psi_{(\theta'\!-\theta+\hat{\theta}, s'\!-s+\hat{s})}\!\left((U^{-1}_{\hat{\theta},\mu^{\hat{s}}}U_{\theta,\mu^s})^{-1}x'\right) 
\\
& \qquad \qquad 
\cdot f_{(\theta'\!, s')}\l(U^{-1}_{\hat{\theta},\mu^{\hat{s}}}x + x'\r)\sigma(x')\sigma(s')\sigma(\theta')
\\
= &\int_{R}\int_{S}\int_{\mathbb{R}^2}\! \mu^{-2(s-\hat{s})} \psi_{\l(\theta'\!-(\theta-\hat{\theta}), s'\!-(s-\hat{s})\r)}\!\left(U^{-1}_{\theta-\hat{\theta},\mu^{s-\hat{s}}}x'\right) \\
& \qquad \qquad
\cdot  f_{(\theta'\!, s')}\l(U^{-1}_{\hat{\theta},\mu^{\hat{s}}}x + x'\r)\sigma(x')\sigma(s')\sigma(\theta')
\\
= & \ [\Psi^{H}\circ f]_{(\theta-\hat{\theta},s-\hat{s})}(U^{-1}_{\hat{\theta},\mu^{\hat{s}}}x)
\\
= & \ \pi_{\hat{\theta},\hat{s}}^{H}[\Psi^{H}\circ f]_{(\theta,s)}(x).
\end{aligned}
\end{equation}

It proves that $\Psi^{H}\circ\pi^{H}_{\hat{\theta},\hat{s}}[f] = \pi_{\hat{\theta},\hat{s}}^{H}[\Psi^{H}\circ f]$.
\end{proof}

\bibliography{egbib.bib}

\end{document}